# A Detailed Chemical Analysis of the Red Giant Orbiting Gaia BH3: From Lithium to Thorium*

Zoe Hackshaw[1], Keith Hawkins[1], and Catherine Manea[2]
[1] Department of Astronomy, The University of Texas at Austin, 2515 Speedway Boulevard, Austin, TX 78712, USA; zoehackshaw@utexas.edu
[2] Department of Physics & Astronomy, University of Utah, Salt Lake City, UT 84112, USA
*Received 2025 October 12; revised 2025 October 20; accepted 2025 October 20; published 2025 December 11*

## Abstract

Preliminary astrometric data from the fourth data release of the Gaia mission revealed a 33 $M_\odot$ dark companion to a metal-poor red giant star, deemed Gaia BH3. This system hosts both the most massive known stellar-origin black hole and the lowest-metallicity star yet discovered in orbit around a black hole. The formation pathway for this peculiar stellar–black hole binary system has yet to be determined, with possible production mechanisms that include isolated binary evolution and dynamical capture. The chemical composition of the stellar companion in Gaia BH3 (hereafter BH3⋆) can help constrain the potential formation mechanisms of this system. Here, we conduct the most comprehensive chemical analysis of BH3⋆ to date using high resolution spectra obtained by the Tull Coudé Spectrograph on the 2.7 m Harlan J. Smith Telescope at McDonald Observatory to constrain potential formation mechanisms. We derived 29 elemental abundances ranging from lithium to thorium and find that BH3⋆ is an $\alpha$-enriched ([$\alpha$/Fe] = 0.41), r-I neutron-capture star ([Eu/Fe] = 0.57). We conclude that BH3⋆ shows no chemical peculiarities (defined as deviations from the expected chemical pattern of an r-I halo red giant) in any elements, which is in alignment with both the dynamical capture and isolated binary evolution formation scenarios. With an upper limit detection on Th, we use the Th/Eu chronometer to place limits on the cosmochronometric age of this system. These observations lay the groundwork for heavy-element chemical analysis for subsequent black hole and low-metallicity stellar binaries that will likely be found in Gaia DR4.

## 1. Introduction

Stellar-mass black holes serve as key laboratories for probing both stellar evolution and compact object formation (J. Strader et al. 2012; M. Morscher et al. 2015; B. P. Abbott et al. 2016; C. Bambi 2025). Dormant black holes in binaries are especially valuable because the luminous companion can help constrain the black hole properties without the observational challenges introduced by accretion-driven emission. If the companion star is metal poor, its spectrum preserves the chemical imprint of early nucleosynthetic events, offering a window into the chemical conditions of the early Universe (C. Sneden et al. 2008; A. Frebel 2018; I. U. Roederer et al. 2018). Studies of such systems can provide critical constraints on the environments in which the first generations of stars formed.

Preliminary Gaia DR4 results revealed a 33 $M_\odot$ dark companion in an 11.6 yr, eccentric ($e = 0.73$) orbit with a metal-poor 0.8 $M_\odot$ red giant star (hereafter BH3⋆; Gaia Collaboration et al. 2024) identified through Gaia astrometry and confirmed with the Radial Velocity Spectrometer (RVS; M. Cropper et al. 2018). This system, deemed Gaia BH3, is located ∼590 pc away in the Galactic halo and shows no detectable emission from the black hole, with only upper limits observed in the radio (L. O. Sjouwerman & J. M. Blanchard 2024), infrared (P. Kervella et al. 2025), ultraviolet (B. Sbarufatti et al. 2025), and X-ray (N. Cappelluti et al. 2024; M. Gilfanov et al. 2024). The lack of electromagnetic radiation across all wavelengths supports the interpretation of the dark companion being a dormant black hole in a binary with a low-metallicity star.

While two other dormant black holes in binaries have been previously discovered with Gaia data (S. Chakrabarti et al. 2023; K. El-Badry et al. 2023a, 2023b; P. Nagarajan et al. 2023; A. Tanikawa et al. 2023), this system is particularly unique. It is the most massive stellar-origin black hole known to date, making it a valuable laboratory for studying hierarchical black hole formation and the possible seeds of intermediate-mass black holes, especially in the low-metallicity regime (e.g., K. Belczynski et al. 2016). This is the lowest metallicity star found in a binary with a black hole, and, consequently, this system can provide constraints on stellar evolution in the early Universe (e.g., J. Merritt et al. 2025).

Another unique aspect of the Gaia BH3 system is its association with the halo stream ED-2, making this the first black hole to be found in a disrupted star cluster (E. Balbinot et al. 2024). ED-2 is a cold (low velocity dispersion) stream currently passing through the solar neighborhood on a highly retrograde orbit (E. Balbinot et al. 2023; E. Dodd et al. 2023). E. Balbinot et al. (2024) proposed that the ED-2 progenitor was a star cluster due to the low dispersion of metallicities among ED-2 stars. E. Dodd et al. (2025) reported 22 chemical abundances of BH3⋆ and other ED-2 stars using the Ultraviolet and Visual Echelle Spectrograph (H. Dekker et al. 2000) and supported the conclusion that ED-2 originated as an ancient star cluster. We expand on this work by reporting more heavy-element abundances with higher signal-to-noise ratio (SNR) spectra of BH3⋆, using these elements to place a

* This paper includes data taken with the Harlan J. Smith 2.7 m telescope at The McDonald Observatory of The University of Texas at Austin.

cosmocrhonometric age limit on this star, and confirming the lack of chemical peculiarities in BH3⋆.

This black hole is the first of its kind located in a disrupted star cluster, and as such the formation pathway of this kind of system is still an open question. There are two main hypothesized formation pathways for Gaia BH3: isolated binary evolution or dynamical capture. In the isolated binary evolution scenario, the two stars would have been born together and remain bound even after the event that created the black hole. G. Iorio et al. (2024), using models of evolved stellar binaries, concluded that Gaia BH3 could plausibly have formed through isolated binary evolution, but only if the natal kick was mild enough to reproduce the orbital parameters found in Gaia Collaboration et al. (2024). However, K. El-Badry (2024) argued that the black hole would have formed with a significant natal kick that likely would have ejected the system from ED-2. Given Gaia BH3's association with a disrupted star cluster, the density of stars could have caused BH3⋆ to be dynamically captured in the orbit of the black hole. D. Marín Pina et al. (2024) analysed star-by-star simulations of globular clusters (GCs) and reproduced BH3-like systems that dynamically formed with a black hole and previously unbound star.

In both the isolated binary evolution scenario and the dynamical capture scenario, chemical peculiarities of BH3⋆ are predicted to be unlikely. In the dynamical capture scenario, BH3⋆ would have no natal association with the black hole and would look chemically similar to other ED-2 or halo stars. In the isolated binary evolution scenario, there is the possibility of ejecta or mass being accreted onto the secondary star in the event that the primary star goes supernova (SN; i.e., P. B. Marks & M. J. Sarna 1998; J. I. González Hernández et al. 2008; Z.-W. Liu et al. 2015; A. Batta et al. 2017). However, K. El-Badry (2024) found that the detection of pollution in BH3 is improbable due to the separation of the two objects and the deep convective envelope of the red giant. Chemical abnormalities may be detectable if the companion star was on the main sequence, yet the dredge-up that occurs in giant stars would likely erase all observable signatures of chemical abnormalities (G. Iorio et al. 2024; A. Rosselli-Calderon et al. 2025).

Here, we conduct the most comprehensive analysis of elemental abundances to confirm the absence of chemical peculiarities in BH3⋆ using the highest SNR spectrum of this star to date. We present the most complete elemental abundance analysis for this system, spanning 29 elements from light elements (e.g., Li) through heavy neutron-capture elements (Th).

Throughout this work, we aim to confirm the lack of chemical peculiarity in BH3⋆ and also provide new abundances for some of the heaviest elements. In Section 2, we describe the observations and data of BH3⋆. In Section 3, we explain the methods used to derive stellar parameters and chemical abundances. Our results are in Section 4, including the stellar parameters (Section 4.1), the chemical abundances (Section 4.2), and a discussion on the nuclear cosmochronometric age (Section 4.3). We conclude our work in Section 5.

## 2. Data

To accurately characterize elemental abundances and any chemical peculiarities of BH3⋆, we need a spectrum with high enough SNR per resolution element (SNR $\gtrsim$ 150), especially in the near-UV optical, to derive the heaviest *r*-process elemental abundances (e.g., I. U. Roederer et al. 2014a). To achieve this, we obtained 44 hr of observations of Gaia DR3 4318465066420528000 (BH3⋆) spanning a range of 11 nights from mid- to late-2024 using the optical ($3700 < \lambda < 9000$ Å) Tull Coudé spectrograph (TS; R. G. Tull et al. 1995) on the 2.7 m Harlan J. Smith Telescope at McDonald Observatory. We obtained 87 1800 s exposures using the TS23 setup with Slit 4, a setup to achieve a resolving power of $R \sim 60{,}000$ (where the resolving power is defined as $R = \lambda/\Delta\lambda$) and yielding the highest SNR spectrum of this star to date (SNR $\sim$ 200). With a resolving power of ∼60,000 that is relatively constant order-to-order along the two-dimensional echelle spectrum, we can denote the resolution ($\Delta\lambda$) to be ∼0.0667 Å at the λ4000 region, ∼0.10 Å at the λ6000 region, and ∼0.133 Å at the λ8000 region.

The raw TS data were processed using the Tull Coudé Spectrograph Data Reduction Pipeline (`TSDRP`[3]). This pipeline performs essential calibration and extraction steps, including bias subtraction, trace identification, scattered light subtraction, wavelength calibration, flat-field correction, cosmic ray rejection, and spectral extraction for each spectral order. Additionally, `TSDRP` provides deblazing, continuum normalization, and order combination to produce a single, fully processed one-dimensional spectrum.

For each exposure we determined the radial velocity (RV) using `iSpec`, a Python code that applies a cross-correlation technique to a template spectrum and an observed spectrum (S. Blanco-Cuaresma et al. 2014; S. Blanco-Cuaresma 2019). As such, we determined the RVs for each exposure and combined them into three epochs and plotted them on the RV curve of Gaia BH3 in Figure 1, similar to Figure 3 of Gaia Collaboration et al. (2024). We show that our RV epochs align with the predicted RV solution for Gaia BH3 using the orbital parameters reported by Gaia Collaboration et al. (2024). With our RVs, we placed all of our exposures at rest and interpolated them onto a common wavelength grid. We median stacked the spectra to get one continuum-normalized and RV-corrected spectrum which we used for our analysis. To illustrate the high quality of this spectrum, we show four representative spectral regions (including Ca H&K, H$\beta$, Mg I triplet, and CH *G* band) in Figure 2.

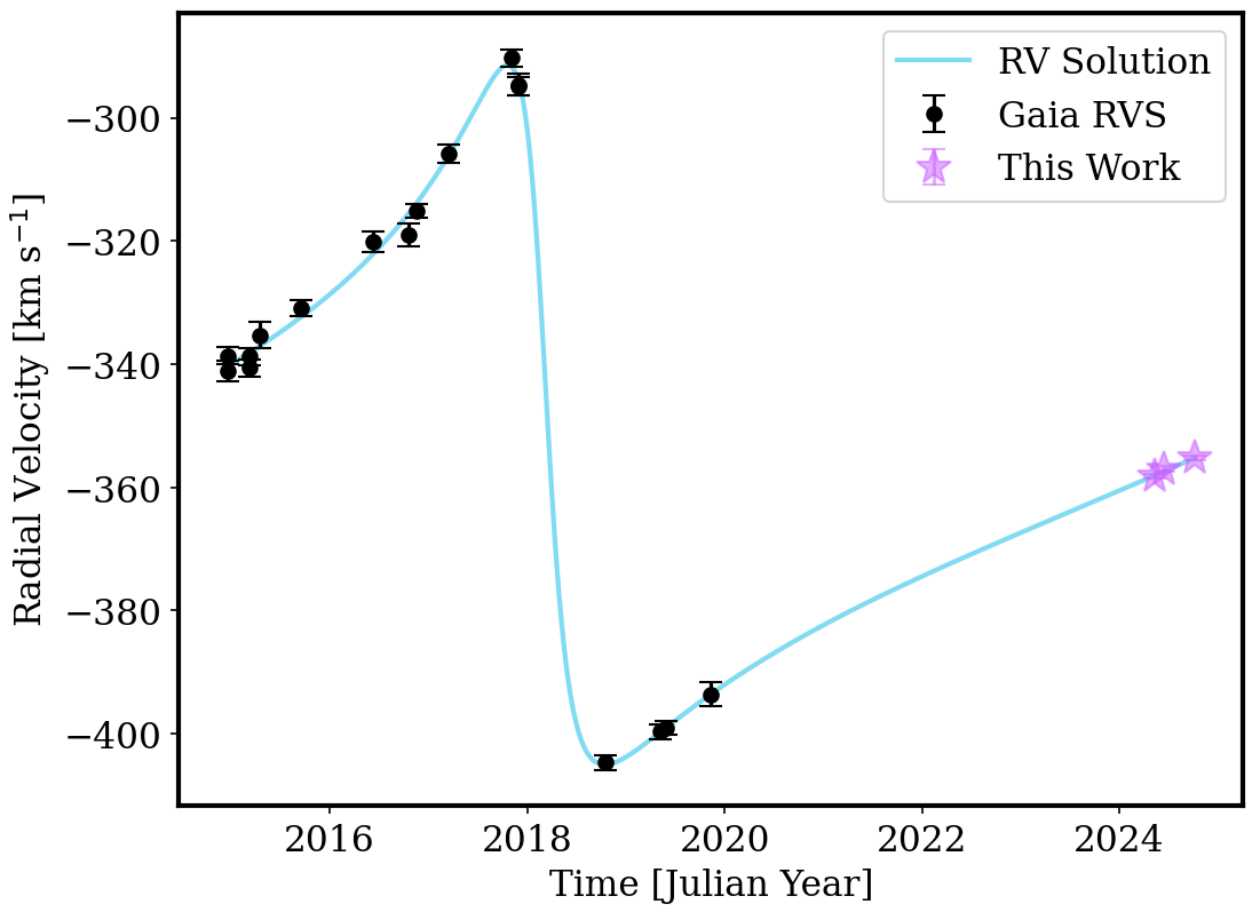

**Figure 1.** Similar to Figure 3 from Gaia Collaboration et al. (2024), the radial velocity evolution of Gaia BH3. The blue line is the radial velocity evolution predicted by the Gaia combined binary model and the Gaia RVS epoch data are shown in black. We combine our RV points spanning 11 nights into three epochs (purple stars).

[3] https://github.com/grzeimann/TSDRP

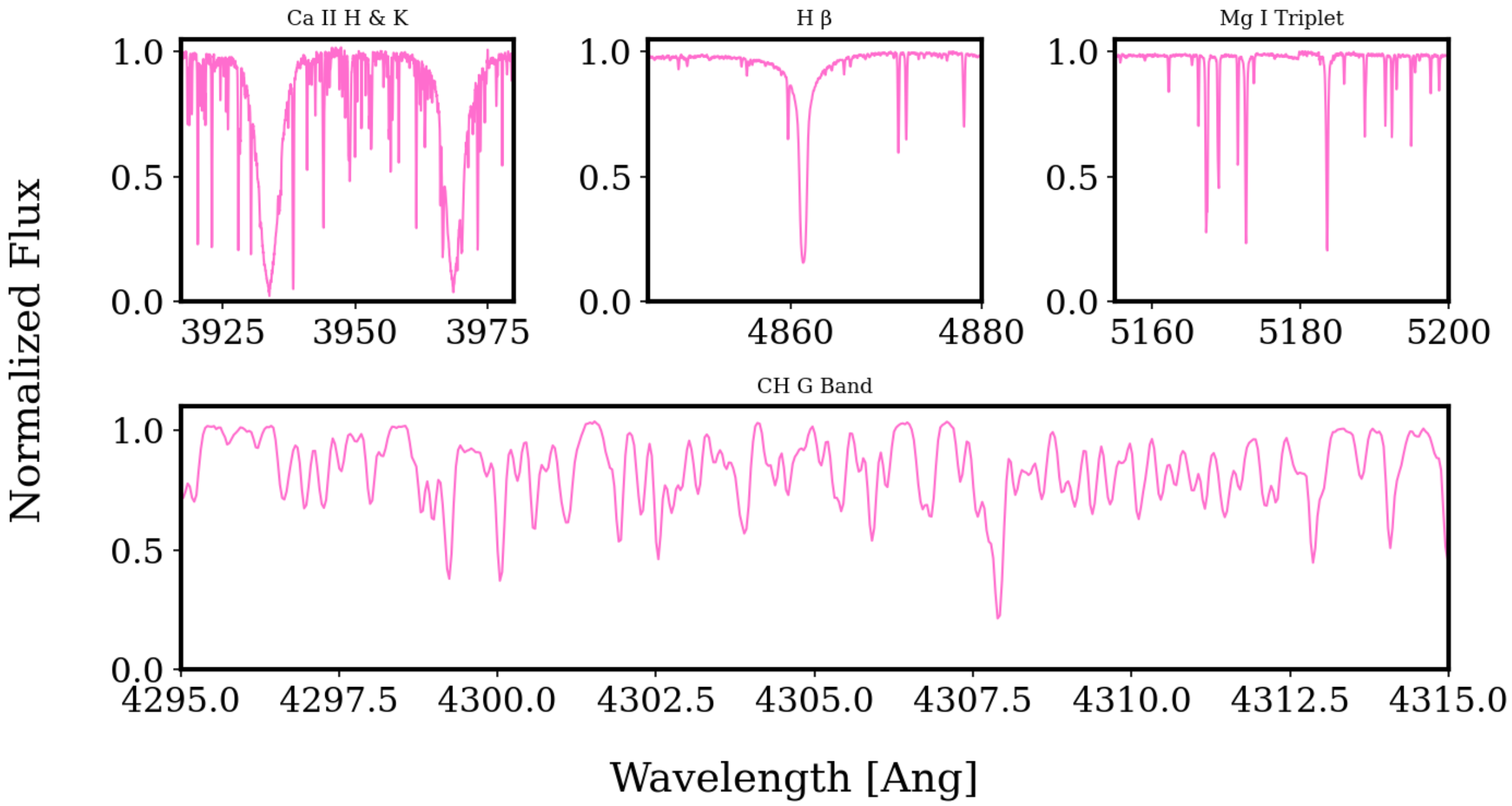

**Figure 2.** Four representative regions of the continuum-normalized and RV-corrected spectrum of BH3⋆ including Ca II H & K, H$\beta$, the Mg I triplet, and the CH *G* band.

For the determination of BH3⋆'s stellar parameters, we use the *G*-band magnitude from Gaia Collaboration et al. (2022) and the $K_s$ magnitude from the Two Micron All Sky Survey (2MASS; R. M. Cutri et al. 2003). For the surface gravity determination, we adopted a mass of 0.76 $M_\odot$ reported by Gaia Collaboration et al. (2024). This system's low Galactic latitude of $b \simeq -3.5°$ requires that we correct for dust extinction because it is in a relatively high extincted area of the Galaxy (J. L. Vergely et al. 2022; Gaia Collaboration et al. 2024). To apply our dust correction in the following section, we use the the Bayestar three-dimensional dust map (G. M. Green et al. 2019) and adopt a distance of 590.6 pc from Gaia Collaboration et al. (2024). With these data, we can determine the stellar parameters and chemical abundances of BH3⋆.

## 3. Methods

To derive the chemical abundances of BH3⋆ and search for any peculiarities, we first need to find the stellar parameters of BH3⋆ to ensure we are using the suitable model atmospheres in our abundance analysis. One typical way of spectroscopically deriving $T_{\rm eff}$ and log $g$ is by using the Fe I and Fe II excitation potential and ionization balance. However, this method would likely yield inaccurate results due to the strong nonlocal thermodynamic equilibrium (NLTE) effects in Fe for metal-poor red giant stars (e.g, C. I. Short & P. H. Hauschildt 2003; M. Fabrizio et al. 2012; R. Ezzeddine et al. 2013, 2017; M. Bergemann & T. Nordlander 2014). For completeness, we report the stellar parameters derived through LTE excitation and ionization balance as $T_{\rm eff}$ = $4904 \pm 12$ K, log $g$ = $1.89 \pm 0.38$ dex, and [Fe/H] = $-2.69 \pm 0.09$. In general, these stellar parameters resulted in systematically higher [X/Fe] abundances when compared to the abundances derived with our photometric stellar parameters.

We avoid using the excitation and ionization balance stellar parameters due to the NLTE effects and instead obtain more reliable stellar parameters for BH3⋆ using the $G - K_s$ color and the $K_s$ magnitude. Photometric temperatures and surface gravities can reduce the discrepancies and provide more accurate stellar parameters (e.g., L. Casagrande et al. 2010; K. Lind et al. 2012; A. Frebel et al. 2013). We derived the effective temperature and surface gravity of BH3⋆ following the procedures outlined in A. Mucciarelli et al. (2021) and L. Casagrande & D. A. VandenBerg (2014), respectively. We derived effective temperatures using the Gaia EDR3-2MASS color-$T_{\rm eff}$ transformations of A. Mucciarelli et al. (2021). These relate $T_{\rm eff}$ to Gaia *G*-band magnitude and 2MASS $K_s$ colors through metallicity-dependent polynomial fits calibrated on infrared flux method (IRFM) temperatures. IRFM temperatures are estimated by comparing the ratio of a star's bolometric flux to its infrared monochromatic flux with that predicted by stellar atmosphere models (i.e., M. S. Bessell et al. 1998). For BH3⋆, we adopted the $G - K_s$ calibrations, which have the smallest scatter ($\sigma_{T_{\rm eff}} \lesssim$ 50 K) and are well-suited for low-metallicity stars. To account for reddening, $E(B - V)$ was calculated using `dustmaps` (G. Green 2018) assuming the Bayestar three-dimensional dust map (G. M. Green et al. 2019). With this method, we find an effective temperature of 5416 ± 84 K.

Following L. Casagrande et al. (2014), we estimate the surface gravity of BH3⋆ by combining synthetic bolometric corrections (BCs) with the $K_s$ magnitude and parallax. Their method leverages the MARCS model atmosphere fluxes to generate synthetic colors and BCs across a grid of stellar parameters, making this method more accurate when deriving surface gravities for metal-poor stars. We combined the $K_s$ magnitude with the BC to derive a bolometric magnitude which was then converted to luminosity. We used this luminosity, our derived $T_{\rm eff}$, and the mass reported in Gaia

**Table 1**
Stellar Parameters of BH3⋆

| Parameter | This Work | Gaia+24 | Dodd+25 |
|---|---|---|---|
| $T_{\rm eff}$ [K] | 5416 ± 84 | 5212 ± 80 | 5445 |
| log $g$ | 3.00 ± 0.14 | 2.929 ± 0.003 | 3.037 |
| [Fe/H] | −2.27 ± 0.24 | −2.56 ± 0.11 | −2.34 |
| [$\alpha$/Fe] | 0.42 ± 0.17 | 0.43 ± 0.12 | 0.40 |
| $\xi$ [km s$^{-1}$] | 1.54 ± 0.11 | ⋯ | 1.60 |

Collaboration et al. (2024) to find a surface gravity of log $g = 3.00 \pm 0.14$ using the Stefan–Boltzmann equation (e.g., D. F. Gray 2005). We report the stellar parameters derived using $G - K_s$ color and the $K_s$ magnitude in Table 1 and adopt these as the BH3⋆ stellar parameters for the rest of this work.

We fixed our photometrically derived effective temperature and surface gravity in the Brussels Automatic Code for Characterizing High Accuracy Spectra (BACCHUS) to determine microturbulence, [Fe/H], and our elemental abundances. BACCHUS (T. Masseron et al. 2016) is an automated framework for deriving stellar parameters and chemical abundances. It performs spectral synthesis of narrow regions around each line of interest using the radiative transfer code `TURBOSPECTRUM` (B. Plez 2012) in combination with MARCS model atmospheres (B. Gustafsson et al. 2008). We refer the reader to K. Hawkins et al. (2016), C. R. Hayes et al. (2022), and the BACCHUS manual[4] for a comprehensive description of the tool. In summation, BACCHUS interpolates across the MARCS model atmosphere grid to choose a model atmosphere at the correct $T_{\rm eff}$, log $g$, [M/H], and $\xi$ to then synthesize the line of interest. BACCHUS scales the solar abundance of the line to the model metallicity and synthesizes a line with this abundance as well as four other lines with −0.6, −0.3, +0.3, and +0.6 dex departures from the scaled solar abundance.

For the line information, we used the solar abundances from M. Asplund et al. (2009) and adopted version 5 of the Gaia-ESO line list (U. Heiter et al. 2021) for atomic data, supplemented with entries from the VALD database for lines with $\lambda \leqslant 4200$ Å (F. Kupka & T. A. Ryabchikova 1999; T. Ryabchikova et al. 2015). We use the following sources for our molecular data: CH (T. Masseron et al. 2014) and C2, CN, OH, and MgH (Masseron, private communication).

We did a line-by-line analysis with BACCHUS and only considered lines where all `flags = 1`, ensuring that no lines were upper limits, extrapolations from the synthesis, or extreme deviations from the observed spectrum and the synthesis. From BACCHUS, we adopted the adopted the $\chi^2$ abundances.

BACCHUS then fits a polynomial to the $\chi^2$ versus log($A$) trend and adopts the abundance that minimizes this trend. Following this method, the chemical abundances for BH3⋆ are found in Table 2, outlining the species, the number of lines (N), the absolute abundance of the species (log $\epsilon$), the standard error of the species ($\sigma_{\rm SE}$), [X/H], [X/Fe], and the total uncertainty on [X/Fe] ($\sigma_{\rm [X/Fe]}$).

To estimate the uncertainties on the elemental abundances, we first found the standard error on the mean (defined as $\sigma_{SE} = \sigma/\sqrt{N}$) of the absolute elemental abundances. For

[4] BACCHUS Manual: https://drive.google.com/file/d/1VShSwA5M21q2pSSxLxc9AnoA19ixu-eV/view.

**Table 2**
The Full Suite of Stellar Abundances for BH3⋆, Outlining the Species (Element), the Number of Lines (N), the Absolute Abundance of the Species (log $\epsilon$), the Standard Error of the Species ($\sigma_{\rm SE}$), [X/H], [X/Fe], and the Total Uncertainty on [X/Fe] ($\sigma_{\rm [X/Fe]}$)

| Element | $N$ | log $\epsilon$ | $\sigma_{\rm SE}$ | [X/H] | [X/Fe] | $\sigma_{\rm [X/Fe]}$ |
|---|---|---|---|---|---|---|
| Li I | 1 | 1.37 | 0.10 | +0.32 | +2.59 | 0.16 |
| C I | 1 | 6.42 | 0.10 | −2.01 | +0.27 | 0.26 |
| O I | 1 | 6.98 | 0.10 | −1.71 | +0.56 | 0.11 |
| Na I | 2 | 3.96 | 0.16 | −2.28 | −0.01 | 0.18 |
| Mg I | 3 | 5.72 | 0.05 | −1.88 | +0.39 | 0.12 |
| Si I | 4 | 5.60 | 0.08 | −1.91 | +0.36 | 0.10 |
| Si II | 1 | 5.46 | 0.10 | −2.05 | +0.22 | 0.34 |
| Ca I | 15 | 4.39 | 0.02 | −1.95 | +0.32 | 0.09 |
| Sc II | 7 | 0.95 | 0.02 | −2.20 | +0.07 | 0.09 |
| Ti I | 22 | 2.99 | 0.04 | −1.96 | +0.31 | 0.15 |
| Ti II | 11 | 2.96 | 0.04 | −1.99 | +0.28 | 0.07 |
| V II | 1 | 1.89 | 0.10 | −2.04 | +0.23 | 0.17 |
| Cr I | 6 | 3.17 | 0.02 | −2.47 | −0.20 | 0.22 |
| Cr II | 3 | 3.30 | 0.07 | −2.34 | −0.07 | 0.09 |
| Mn I | 1 | 2.64 | 0.10 | −2.79 | −0.52 | 0.16 |
| Fe I | 56 | 5.22 | 0.02 | −2.28 | −0.01 | 0.13 |
| Fe II | 21 | 5.24 | 0.03 | −2.26 | +0.00 | 0.08 |
| Co I | 1 | 2.83 | 0.10 | −2.16 | +0.11 | 0.15 |
| Ni I | 9 | 4.04 | 0.04 | −2.18 | +0.09 | 0.12 |
| Cu I | 1 | 2.11 | 0.10 | −2.08 | +0.19 | 0.14 |
| Zn I | 1 | 2.38 | 0.10 | −2.18 | +0.08 | 0.12 |
| Sr II | 1 | 0.58 | 0.10 | +2.28 | +0.02 | 0.12 |
| Y I | 1 | −0.35 | 0.10 | −2.56 | −0.29 | 0.14 |
| Y II | 4 | −0.36 | 0.04 | −2.57 | −0.30 | 0.10 |
| Zr II | 2 | 0.54 | 0.01 | −2.04 | +0.23 | 0.10 |
| Ba II | 3 | −0.21 | 0.04 | −2.39 | −0.12 | 0.13 |
| La II | 3 | −0.90 | 0.06 | −2.00 | +0.26 | 0.11 |
| Ce II | 2 | −0.49 | 0.03 | −2.07 | +0.20 | 0.20 |
| Pr II | 1 | −1.34 | 0.10 | −2.06 | +0.21 | 0.16 |
| Nd II | 4 | −0.38 | 0.04 | −1.80 | +0.47 | 0.16 |
| Sm II | 1 | −0.51 | 0.10 | −1.47 | +0.80 | 0.15 |
| Eu II | 1 | −1.18 | 0.10 | −1.70 | +0.57 | 0.13 |
| Dy II | 1 | −0.50 | 0.10 | −1.60 | +0.67 | 0.14 |
| Th II | 1 | $\lesssim$−2.00 | 0.10 | $\lesssim$−2.02 | $\lesssim$ +0.25 | 0.62 |

**Note.** The last column ($\sigma_{\rm [X/Fe]}$) represents the error on [X/Fe], combining $\sigma_{\rm SE}$ and the sensitivity abundance uncertainties found in Table 3. For species with only one line measured, we conservatively assume $\sigma_{\rm SE}$ = 0.10 dex.

elements with only one line measured, we conservatively assume $\sigma_{\rm SE} = 0.1$ dex. We then combined the standard error on each species with the uncertainties found when varying the stellar parameters by $\sim 1\sigma$ ($T_{\rm eff} \sim 100$ K, log $g \sim 0.15$ dex, and $\xi \sim 0.15$ km s$^{-1}$) and rederiving the abundances, assuming that the errors on log $g$, $T_{\rm eff}$, and $\xi$ are independent of one another. The results of these sensitivity tests (Table 3) are then added in quadrature with the standard error to report the total uncertainty on the abundances in the final column of Table 2.

## 4. Results and Discussion

### 4.1. Stellar Parameters

Using the photometric procedures outlined above, we determined that BH3⋆ is a metal-poor red giant star with an effective temperature of 5416 ± 84 K and a surface gravity of 3.00 ± 0.14 dex. Our derived stellar parameters along with stellar parameters from different studies of BH3⋆ (Gaia Collaboration et al. 2024; E. Dodd et al. 2025) are found in Table 1. Our $T_{\rm eff}$ agrees within error with the temperature found in E. Dodd et al. (2025) and is about ∼200 K higher than the temperature found in Gaia Collaboration et al. (2024). The difference in temperature between this work and Gaia Collaboration et al. (2024) is likely a result of the different dust maps that were chosen when calculating $E(B-V)$. Our surface gravities agree well within error with both E. Dodd et al. (2025) and Gaia Collaboration et al. (2024). With our derived stellar parameters generally agreeing with different works, we continue to use these stellar parameters and now look toward the elemental abundances.

**Table 3**
Sensitivity Tests for Abundance Uncertainties

| Element | $\Delta T_{\rm eff}$ ±100 K | $\Delta \log g$ ±0.15 dex | $\Delta\xi$ ±0.15 km s$^{-1}$ |
|---|---|---|---|
| Li I | ±0.12 | ±0.02 | ±0.02 |
| C I | ±0.12 | ±0.02 | ±0.02 |
| O I | ±0.03 | ±0.01 | ±0.02 |
| Na I | ±0.07 | ±0.01 | ±0.01 |
| Mg I | ±0.10 | ±0.01 | ±0.01 |
| Si I | ±0.04 | ±0.0 | ±0.02 |
| Si II | ±0.23 | ±0.22 | ±0.01 |
| Ca I | ±0.08 | ±0.01 | ±0.01 |
| Sc II | ±0.08 | ±0.01 | ±0.02 |
| Ti I | ±0.14 | ±0.01 | ±0.02 |
| Ti II | ±0.03 | ±0.03 | ±0.01 |
| V II | ±0.13 | ±0.01 | ±0.03 |
| Cr I | ±0.19 | ±0.06 | ±0.05 |
| Cr II | ±0.05 | ±0.02 | ±0.01 |
| Mn I | ±0.10 | ±0.01 | ±0.06 |
| Fe I | ±0.12 | ±0.01 | ±0.03 |
| Fe II | ±0.07 | ±0.01 | ±0.01 |
| Co I | ±0.10 | ±0.01 | ±0.03 |
| Ni I | ±0.10 | ±0.02 | ±0.02 |
| Cu I | ±0.10 | ±0.01 | ±0.03 |
| Zn I | ±0.05 | ±0.02 | ±0.01 |
| Sr II | ±0.05 | ±0.03 | ±0.01 |
| Y I | ±0.09 | ±0.02 | ±0.01 |
| Y II | ±0.09 | ±0.01 | ±0.02 |
| Zr II | ±0.09 | ±0.01 | ±0.03 |
| Ba II | ±0.12 | ±0.02 | ±0.03 |
| La II | ±0.09 | ±0.02 | ±0.03 |
| Ce II | ±0.12 | ±0.13 | ±0.10 |
| Pr II | ±0.12 | ±0.01 | ±0.02 |
| Nd II | ±0.13 | ±0.08 | ±0.01 |
| Sm II | ±0.10 | ±0.02 | ±0.01 |
| Eu II | ±0.08 | ±0.03 | ±0.01 |
| Dy II | ±0.10 | ±0.01 | ±0.01 |
| Th II | ±0.61 | ±0.05 | ±0.01 |

### *4.2. Elemental Abundances*

In this section, we describe the stellar abundances derived for BH3⋆ with the line list outlined in Table 4. We plot the abundances of BH3⋆ along with the dwarf spheroidal (dSph) galaxy Sculptor (D. Geisler et al. 2005), the *r*-process enhanced ultrafaint dwarf (UFD) galaxy Reticulum II (A. P. Ji et al. 2016), the host stellar stream of BH3⋆ ED-2 (E. Dodd et al. 2025), the metal-poor GC M15 (J. S. Sobeck et al. 2011), and MW halo stars (I. U. Roederer et al. 2014b) to illustrate where systems with different chemical evolution histories may lie in these planes. We note that the stellar parameters for the I. U. Roederer et al. (2014b) stars have been updated in S. Mittal & I. U. Roederer (2025) resulting in generally smaller abundances, yet we still plot the I. U. Roederer et al. (2014b) abundances due to the wide range of chemical abundances reported in the initial sample.

#### *4.2.1. Light Elements*

We determined the abundances of the light elements Li, C, and Na. These light element abundances for BH3⋆ are plotted in Figure 3 along with Sculptor, Reticulum II, ED-2, M15, and other halo stars.

The majority of lithium (Li) in the Universe was created during Big Bang nucleosynthesis and the Li abundance of extremely metal-poor unevolved dwarf stars can represent the primordial value (F. Spite & M. Spite 1982). A smaller fraction of Li is thought to be synthesized by the $\nu$-process, novae, and SNe (F. X. Timmes et al. 1995; S. Borisov et al. 2024). The Li abundance for BH3⋆ was derived from the Li I 6707 Å line, yielding [Li I/Fe] = 2.57. The absolute lithium abundance of $A$(Li) = 1.34 places this star below the Spite plateau (F. Spite & M. Spite 1982), as expected for a red giant. At this evolutionary stage, the star has likely begun depleting its surface Li through convective mixing and internal burning (e.g., A. Skumanich 1972; C. Charbonnel & S. C. Balachandran 2000; L. Sbordone et al. 2010; R. Singh et al. 2025). The Spite plateau (purple line) is illustrated in the left panel of Figure 3 by Milky Way halo dwarf (triangle) and giant (square) stars and plotted alongside BH3⋆. We do not see any chemical peculiarities in the Li abundance of BH3⋆ and it is well predicted by previous studies of Li in metal-poor giants, such as C. A. Pilachowski et al. (1993).

Carbon (C) is produced both by nucleosynthesis in massive ($M > 10\ M_\odot$) stars and low- to intermediate- mass asymptotic giant branch (AGB) stars (C. Kobayashi et al. 2020). The carbon abundance of this star was derived using mainly the CH band at ∼4300 Å. We note that there may be complications with using solely the CH $G$ band for the derivation of carbon (i.e., P. Santos-Peral et al. 2025), yet no atomic C features were detected in this star. We find [C I/Fe] = 0.27 ± 0.16, which is below the threshold to classify this star as a carbon-enhanced metal-poor star, typically defined as stars with [C/Fe] > +0.7 (e.g, T. C. Beers et al. 1992; T. C. Beers & N. Christlieb 2005; N. Christlieb et al. 2008). We see no chemical peculiarities in the C abundance of BH3⋆, shown in the middle panel of Figure 3.

Sodium (Na) is mainly produced during carbon-burning in massive stars (S. E. Woosley & T. A. Weaver 1995; C. Kobayashi et al. 2006; K. Nomoto et al. 2013) and enriches the interstellar medium (ISM) in Type II SNe (F. X. Timmes et al. 1995; K. A. Venn et al. 2004). Na can also be synthesized during the third dredge-up in AGB stars (N. Mowlavi 1999). We derived a sodium abundance of [Na I/Fe] = 0.0 ± 0.18 using two Na I lines at 8183.3 Å and 8194.8 Å. We see no chemical peculiarities in the Na abundance of BH3⋆, shown in the right panel of Figure 3.

#### *4.2.2. α Elements*

We found a global α abundance of [α/Fe] = 0.41 based on the α element abundances of O, Mg, Si, and Ca. The abundances of α elements for BH3⋆ are plotted in Figure 4 along with Sculptor, Reticulum II, ED-2, M15, and other halo stars.

The [α/Fe] of BH3⋆ indicates that this star is α-enriched, which is typical of older halo stars. This enrichment reflects

**Table 4**
The Atomic Line Information for the Lines Used in this Analysis

| λ (Å) | El./Ion | χ (eV) | log *gf* | log ϵ(*x*) (dex) | λ (Å) | El./Ion | χ (eV) | log *gf* | log ϵ(*x*) (dex) | λ (Å) | El./Ion | χ (eV) | log *gf* | log ϵ(*x*) (dex) |
|---|---|---|---|---|---|---|---|---|---|---|---|---|---|---|
| 6707.915 | Li I | 0.0 | −0.303 | 1.368 | 5147.478 | Ti I | 0.0 | −1.94 | 3.157 | 5151.911 | Fe I | 1.011 | −3.322 | 5.391 |
| 7771.944 | O I | 9.146 | 0.369 | 6.98 | 5210.384 | Ti I | 0.048 | −0.82 | 3.028 | 5162.272 | Fe I | 4.178 | 0.02 | 5.256 |
| 8183.255 | Na I | 2.102 | 0.237 | 4.193 | 5219.702 | Ti I | 0.021 | −2.22 | 3.138 | 5171.596 | Fe I | 1.485 | −1.721 | 5.294 |
| 8194.79 | Na I | 2.104 | −0.462 | 3.728 | 5224.3 | Ti I | 2.134 | 0.13 | 2.764 | 5197.577 | Fe I | 4.956 | −2.078 | 5.163 |
| 4571.096 | Mg I | 0.0 | −5.623 | 5.818 | 5490.697 | Ti I | 2.154 | −5.508 | 2.667 | 5198.711 | Fe I | 2.223 | −2.135 | 5.318 |
| 5528.405 | Mg I | 4.346 | −0.498 | 5.603 | 5490.846 | Ti I | 0.048 | −3.35 | 2.667 | 5216.274 | Fe I | 1.608 | −2.082 | 5.276 |
| 5711.088 | Mg I | 4.346 | −1.724 | 5.743 | 5866.368 | Ti I | 3.305 | −1.055 | 3.111 | 5217.389 | Fe I | 3.211 | −1.074 | 5.148 |
| 5645.613 | Si I | 4.93 | −2.043 | 5.761 | 6064.626 | Ti I | 1.046 | −1.888 | 3.418 | 5225.526 | Fe I | 0.11 | −4.789 | 5.457 |
| 5665.555 | Si I | 4.92 | −1.94 | 5.391 | 6258.102 | Ti I | 1.443 | −0.39 | 3.204 | 5232.94 | Fe I | 2.94 | −0.07 | 5.106 |
| 5708.399 | Si I | 4.954 | −1.37 | 5.469 | 6261.099 | Ti I | 1.43 | −0.53 | 3.114 | 5250.205 | Fe I | 5.012 | −3.248 | 5.442 |
| 5793.073 | Si I | 4.93 | −1.963 | 5.769 | 4798.531 | Ti II | 1.08 | −2.66 | 3.097 | 5281.79 | Fe I | 3.039 | −0.833 | 5.13 |
| 6347.109 | Si II | 8.121 | 0.169 | 5.462 | 4865.61 | Ti II | 1.116 | −2.7 | 2.801 | 5302.3 | Fe I | 3.283 | −0.738 | 5.119 |
| 5261.704 | Ca I | 2.521 | −0.579 | 4.247 | 4874.009 | Ti II | 3.095 | −0.86 | 2.704 | 5307.361 | Fe I | 1.608 | −2.912 | 5.278 |
| 5349.465 | Ca I | 2.709 | −0.31 | 4.326 | 4911.194 | Ti II | 3.124 | −0.64 | 2.817 | 5324.179 | Fe I | 3.211 | −0.108 | 5.039 |
| 5512.98 | Ca I | 2.933 | −0.464 | 4.327 | 5005.167 | Ti II | 1.566 | −2.73 | 3.157 | 5339.929 | Fe I | 3.266 | −0.635 | 5.07 |
| 5581.965 | Ca I | 2.523 | −0.555 | 4.303 | 5129.156 | Ti II | 1.892 | −1.34 | 3.014 | 5367.466 | Fe I | 4.415 | 0.444 | 5.01 |
| 5588.749 | Ca I | 2.526 | 0.358 | 4.242 | 5185.902 | Ti II | 1.893 | −1.41 | 2.915 | 5383.369 | Fe I | 4.313 | 0.645 | 4.951 |
| 5590.114 | Ca I | 2.521 | −0.571 | 4.343 | 5211.53 | Ti II | 2.59 | −1.41 | 2.893 | 5393.167 | Fe I | 3.241 | −0.719 | 5.133 |
| 5857.451 | Ca I | 2.933 | 0.24 | 4.327 | 5336.786 | Ti II | 1.582 | −1.6 | 2.985 | 5405.774 | Fe I | 0.99 | −1.849 | 5.462 |
| 6102.723 | Ca I | 1.879 | −0.85 | 4.435 | 5381.021 | Ti II | 1.566 | −1.97 | 3.028 | 5415.199 | Fe I | 4.387 | 0.643 | 4.935 |
| 6122.217 | Ca I | 1.886 | −0.38 | 4.423 | 5418.768 | Ti II | 1.582 | −2.13 | 3.149 | 5424.068 | Fe I | 4.32 | 0.52 | 5.098 |
| 6162.173 | Ca I | 1.899 | −0.17 | 4.492 | 4875.502 | V II† | 5.468 | −1.408 | 1.892 | 5434.523 | Fe I | 1.011 | −2.121 | 5.439 |
| 6166.439 | Ca I | 2.521 | −1.142 | 4.518 | 5206.023 | Cr I | 0.941 | 0.02 | 3.102 | 5501.465 | Fe I | 0.958 | −3.046 | 5.406 |
| 6169.042 | Ca I | 2.523 | −0.797 | 4.502 | 5247.565 | Cr I | 0.961 | −1.59 | 3.206 | 5506.779 | Fe I | 0.99 | −2.795 | 5.44 |
| 6169.563 | Ca I | 2.526 | −0.478 | 4.353 | 5296.691 | Cr I | 0.983 | −1.36 | 3.224 | 5572.842 | Fe I | 3.397 | −0.289 | 5.099 |
| 6493.781 | Ca I | 2.521 | −0.109 | 4.446 | 5300.745 | Cr I | 0.983 | −2.0 | 3.144 | 5576.089 | Fe I | 3.43 | −0.9 | 5.223 |
| 6499.65 | Ca I | 2.523 | −0.818 | 4.531 | 5345.796 | Cr I | 1.004 | −0.95 | 3.198 | 5586.767 | Fe I | 4.26 | −3.023 | 5.059 |
| 4400.389 | Sc II | 0.605 | −0.536 | 0.954 | 5348.314 | Cr I | 1.004 | −1.21 | 3.155 | 5615.62 | Fe I | 4.191 | −2.941 | 5.016 |
| 5031.021 | Sc II | 1.357 | −0.4 | 0.912 | 4588.199 | Cr II | 4.071 | −0.627 | 3.258 | 6065.485 | Fe I | 4.956 | −3.471 | 5.253 |
| 5526.79 | Sc II | 1.768 | 0.024 | 0.837 | 4848.235 | Cr II | 3.864 | −1.18 | 3.468 | 6136.615 | Fe I | 2.453 | −1.402 | 5.263 |
| 5641.001 | Sc II | 1.5 | −1.131 | 1.016 | 5409.798 | Cr II | 12.176 | −6.443 | 3.168 | 6219.313 | Fe I | 5.458 | −5.082 | 5.366 |
| 5657.896 | Sc II | 1.507 | −0.603 | 0.951 | 4823.495 | Mn I† | 2.319 | −2.545 | 2.638 | 6246.318 | Fe I | 3.603 | −0.771 | 5.123 |
| 5667.149 | Sc II | 1.5 | −1.309 | 0.978 | 4872.083 | Fe I | 4.593 | −2.403 | 5.07 | 6301.5 | Fe I | 3.654 | −0.72 | 5.068 |
| 5684.202 | Sc II | 1.507 | −1.074 | 0.995 | 4891.492 | Fe I | 2.851 | −0.111 | 5.063 | 6335.33 | Fe I | 2.198 | −2.177 | 5.253 |
| 4533.239 | Ti I | 0.848 | 0.54 | 2.626 | 4920.502 | Fe I | 2.833 | 0.068 | 5.133 | 6336.823 | Fe I | 3.686 | −0.852 | 5.205 |
| 4534.776 | Ti I | 0.836 | 0.35 | 2.909 | 4923.908 | Fe I | 4.559 | −1.242 | 5.032 | 6393.601 | Fe I | 2.433 | −1.452 | 5.2 |
| 4548.764 | Ti I | 0.826 | −0.28 | 2.941 | 4924.77 | Fe I | 2.279 | −2.216 | 5.314 | 6411.648 | Fe I | 3.654 | −0.596 | 5.116 |
| 4913.613 | Ti I | 1.873 | 0.22 | 3.074 | 4939.687 | Fe I | 0.859 | −3.336 | 5.379 | 6494.98 | Fe I | 2.404 | −1.268 | 5.578 |
| 4981.73 | Ti I | 0.848 | 0.57 | 2.915 | 4946.387 | Fe I | 3.368 | −1.11 | 5.201 | 6546.238 | Fe I | 2.759 | −1.536 | 5.178 |
| 4997.097 | Ti I | 0.0 | −2.07 | 3.105 | 4994.129 | Fe I | 0.915 | −3.058 | 5.395 | 6592.913 | Fe I | 2.728 | −1.473 | 5.181 |
| 4999.503 | Ti I | 0.826 | 0.32 | 2.935 | 5018.413 | Fe I | 3.237 | −3.21 | 4.986 | 6593.87 | Fe I | 2.433 | −2.42 | 5.469 |
| 5009.645 | Ti I | 0.021 | −2.2 | 3.215 | 5049.82 | Fe I | 2.279 | −1.348 | 5.221 | 6750.151 | Fe I | 2.424 | −2.618 | 5.366 |
| 5016.161 | Ti I | 0.848 | −0.48 | 2.864 | 5060.079 | Fe I | 0.0 | −5.431 | 5.407 | 4416.818 | Fe II | 2.778 | −2.41 | 4.969 |
| 5024.844 | Ti I | 0.818 | −0.53 | 3.077 | 5083.338 | Fe I | 0.958 | −2.939 | 5.359 | 4508.28 | Fe II | 2.856 | −2.25 | 5.049 |
| 5039.941 | Ti I | 2.333 | −2.427 | 3.026 | 5127.359 | Fe I | 0.915 | −3.306 | 5.402 | 4520.218 | Fe II | 2.807 | −2.6 | 5.202 |
| 5145.46 | Ti I | 1.46 | −0.54 | 2.742 | 5141.739 | Fe I | 2.424 | −1.978 | 4.987 | 4576.333 | Fe II | 2.844 | −2.92 | 5.132 |
| 4938.801 | Fe II | 12.942 | −4.103 | 5.159 | 4873.438 | Ni I | 3.699 | −0.38 | 3.857 | 4317.309 | Zr II | 0.713 | −1.45 | 0.554 |

**Table 4**
(Continued)

| $\lambda$ (Å) | El./Ion | $\chi$ (eV) | log $gf$ | log $\epsilon(x)$ (dex) | $\lambda$ (Å) | El./Ion | $\chi$ (eV) | log $gf$ | log $\epsilon(x)$ (dex) | $\lambda$ (Å) | El./Ion | $\chi$ (eV) | log $gf$ | log $\epsilon(x)$ (dex) |
|---|---|---|---|---|---|---|---|---|---|---|---|---|---|---|
| 4993.35 | Fe II | 2.807 | −3.684 | 5.297 | 4976.326 | Ni I | 1.676 | −3.0 | 4.081 | 5853.676 | Ba II† | 0.604 | −1.965 | −0.274 |
| 5068.776 | Fe II | 11.681 | −5.504 | 5.153 | 5003.741 | Ni I | 1.676 | −3.07 | 4.159 | 6141.709 | Ba II† | 0.704 | −0.395 | −0.245 |
| 5194.897 | Fe II | 13.471 | 0.501 | 5.273 | 5035.357 | Ni I | 3.635 | 0.29 | 3.808 | 6496.9 | Ba II† | 0.604 | −0.503 | −0.111 |
| 5234.623 | Fe II | 3.221 | −2.18 | 5.18 | 5476.903 | Ni I | 1.826 | −0.78 | 4.041 | 4920.98 | La II | 0.126 | −0.58 | −0.878 |
| 5264.802 | Fe II | 3.23 | −3.13 | 5.289 | 5587.858 | Ni I | 1.935 | −2.39 | 4.061 | 4921.775 | La II | 0.244 | −0.45 | −1.042 |
| 5284.103 | Fe II | 2.891 | −3.195 | 5.31 | 6176.807 | Ni I | 4.088 | −0.26 | 4.101 | 5114.56 | La II | 0.235 | −1.03 | −0.794 |
| 5316.609 | Fe II | 3.153 | −1.87 | 5.156 | 6643.63 | Ni I | 1.676 | −2.22 | 4.1 | 4539.745 | Ce II | 0.328 | −0.08 | −0.536 |
| 5325.552 | Fe II | 3.221 | −3.16 | 5.238 | 6767.772 | Ni I | 1.826 | −2.14 | 4.113 | 4562.359 | Ce II | 0.478 | 0.21 | −0.441 |
| 5425.249 | Fe II | 3.199 | −3.22 | 5.301 | 5153.227 | Cu I | 3.786 | −0.023 | 2.107 | 4408.81 | Pr II | 0.0 | 0.179 | −1.341 |
| 5534.893 | Fe II | 10.545 | −0.44 | 5.331 | 4810.528 | Zn I | 4.078 | −0.16 | 2.375 | 4446.38 | Nd II | 0.205 | −0.35 | −0.275 |
| 5569.592 | Fe II | 8.23 | −4.19 | 5.116 | 4305.443 | Sr II | 3.04 | −0.136 | 0.015 | 4462.98 | Nd II | 0.559 | 0.04 | −0.478 |
| 6247.576 | Fe II | 5.956 | −4.827 | 5.323 | 4900.085 | Y I | 1.398 | −0.334 | −0.352 | 4959.099 | Nd II† | 0.064 | −0.8 | −0.389 |
| 6252.625 | Fe II | 11.093 | −3.044 | 5.302 | 4398.01 | Y II | 0.13 | −0.895 | −0.313 | 5255.502 | Nd II† | 0.205 | −0.67 | −0.396 |
| 6358.701 | Fe II | 11.746 | −4.404 | 5.548 | 4883.682 | Y II | 1.084 | 0.19 | −0.281 | 4577.69 | Sm II | 0.248 | −0.65 | −0.512 |
| 6456.379 | Fe II | 3.903 | −2.185 | 5.338 | 5087.416 | Y II | 1.084 | −0.16 | −0.407 | 4129.717 | Eu II† | 0.0 | −1.48 | −1.18 |
| 6516.077 | Fe II | 2.891 | −3.31 | 5.275 | 5200.406 | Y II | 0.992 | −0.47 | −0.455 | 4077.966 | Dy II | 0.103 | −0.04 | −0.5 |
| 5483.31 | Co I† | 1.71 | −2.04 | 2.833 | 4208.98 | Zr II | 0.713 | −0.51 | 0.522 | 4019.129 | Th II | 0.0 | −0.228 | $\lesssim$-2.0 |

**Note.** The atomic line information for the lines used in this analysis, sourced from version 5 of the *Gaia*-ESO linelist \citep{heiter2021} for lines with $\lambda > 4200$Åand the VALD database for lines with $\lambda < 4200$Å \citep{Kupka1999,Ryabchikova2015}.The columns include wavelength ($\lambda$), the element and ion (El./Ion), the excitation potential ($\chi$), the logarithm product of the statistical weight and oscillator strength (log $gf$), and the absolute abundance derived from the line (log $\epsilon$(x)). Lines denoted with † have hyperfine structure (HFS) components, yet only one transition is shown for simplicity. The full linelist including HFS data is available upon request.

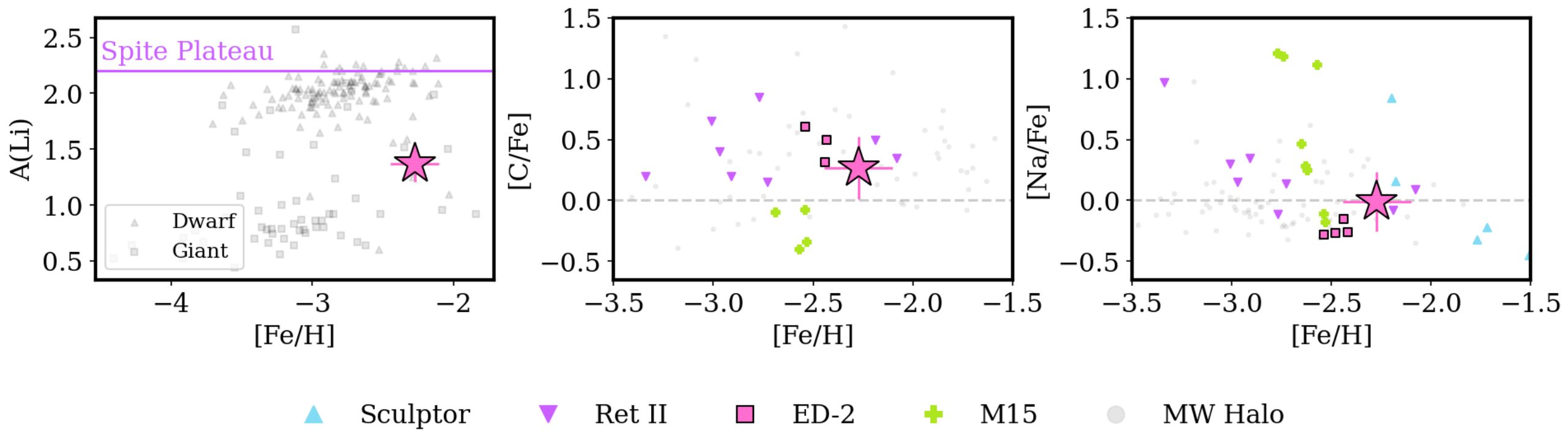

**Figure 3.** A multipanel plot showing the Li (left), C (middle), and Na (right) abundances of BH3⋆ (pink star) compared to the dSph Sculptor (blue triangles; D. Geisler et al. 2005), the UFD Reticulum II (upside-down purple triangles; A. P. Ji et al. 2016), BH3⋆'s host halo stream ED-2 (pink squares; E. Dodd et al. 2025), the GC M15 (green crosses; J. S. Sobeck et al. 2011), and other MW halo stars (gray circles; I. U. Roederer et al. 2014b). In all of the panels, [X/Fe] is on the $y$-axis, with the exception of the first panel which shows lithium in terms of absolute abundance. The dashed line at 0 represents the solar abundance of [X/Fe]. The left panel shows the Spite plateau annotated in purple, illustrating that BH3⋆ lies well below the Spite plateau at $A$(Li) $\sim$ 2.2 dex (F. Spite & M. Spite 1982). We separate dwarf stars (triangles) and giant stars (squares) in the left panel to show that metal-poor dwarf stars make up the Spite plateau and giant stars have largely undergone the burning of their lithium.

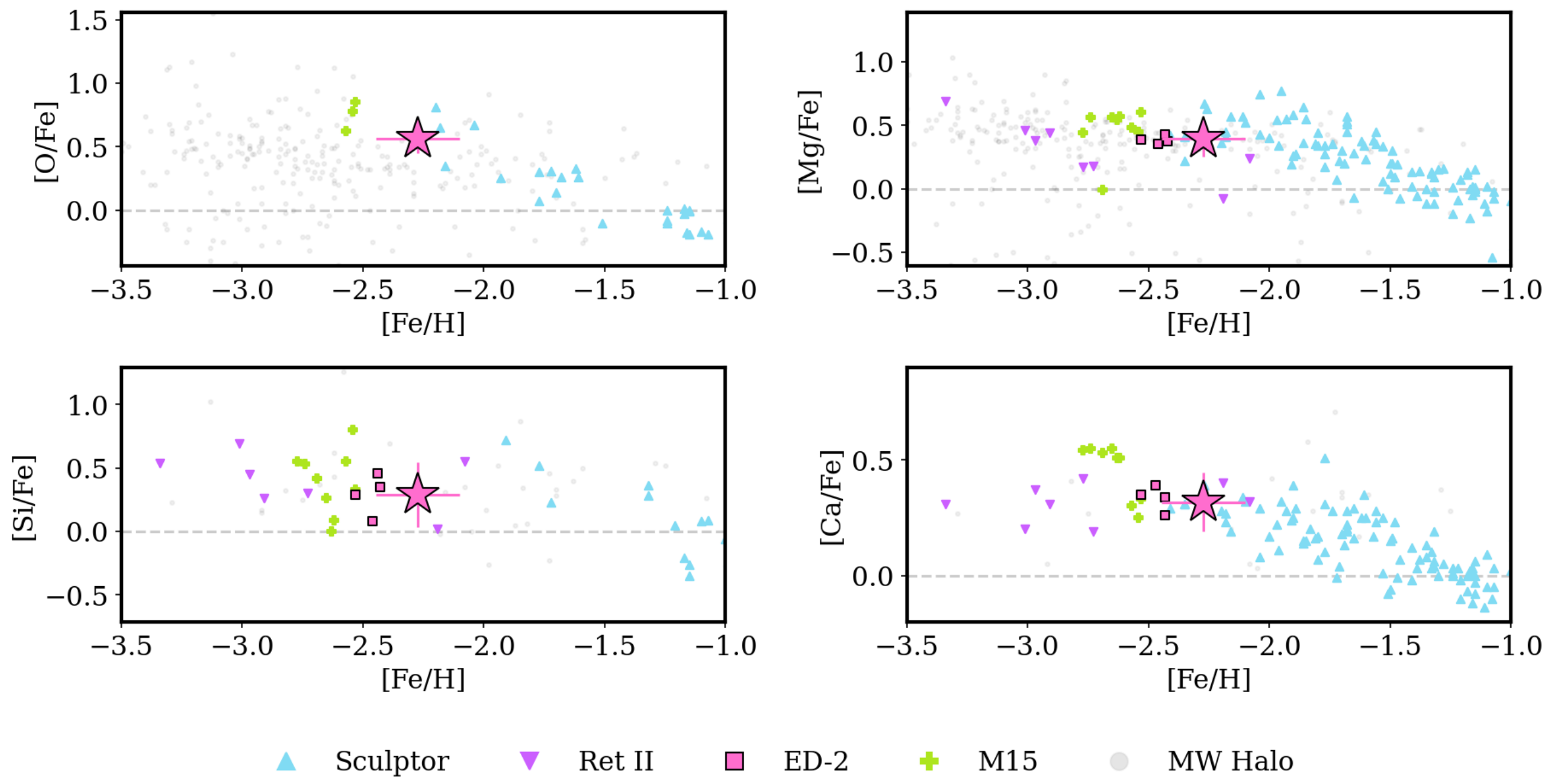

**Figure 4.** Similar to Figure 3 but for $\alpha$-elements.

early chemical evolution, when Type II core-collapse SNe dominated and efficiently enriched the ISM with $\alpha$-elements (e.g., O, Mg, Si, S, and Ca) after being synthesized during the triple-alpha process in massive stars (F. X. Timmes et al. 1995; F.-K. Thielemann et al. 1996; C. Kobayashi et al. 2006; A. Arcones & F.-K. Thielemann 2023). Over time, as Type Ia SNe began to contribute more significantly, the increased production of Fe-peak elements (e.g., Fe, Cr, Mn, Co, Ni) will lead to a decline in [$\alpha$/Fe] ratios in younger stellar populations (e.g., B. M. Tinsley 1980; L. Greggio & A. Renzini 1983; S. E. Woosley & R. D. Hoffman 1992; F. Matteucci & S. Recchi 2001; O. A. Gonzalez et al. 2011). Due to ED-2 originating from a disrupted ancient star cluster, it likely did not have time for Type Ia SNe to dominate, causing ED-2 stars to be $\alpha$ enriched.

We find an oxygen abundance of [O I/Fe] $= 0.56 \pm 0.11$ with one line in the O I triplet at 7771.94 Å. The two other lines in the O I triplet were not selected because the lines exhibit distorted or asymmetric cores that prevent reliable fitting, and thus did not fit our selection criteria of satisfactory BACCHUS flags. The O abundance is shown in the top left panel of Figure 4 with no obvious chemical peculiarities.

We derived a magnesium abundance of [Mg I/Fe] $= 0.39 \pm 0.12$ using three magnesium lines at 4571, 5528, and 5711 Å. We show the Mg abundance in the top right panel of Figure 4, with no obvious chemical peculiarities. The Mg abundance of BH3⋆ generally agrees with the Mg abundance of other ED-2 stars and displays no chemical peculiarities.

We report two species of silicon (Si), with the abundances [Si I/Fe] $= 0.36 \pm 0.10$ and [Si II/Fe] $= 0.22 \pm 0.34$. The overall Si abundance of BH3⋆ is shown in the bottom left panel of Figure 4, with an Si abundance that is generally consistent with other ED-2 stars. We find a calcium abundance of [Ca/Fe] $= 0.32 \pm 0.09$ consistent with other ED-2 stars, shown in the bottom right panel of Figure 4. There are no chemical peculiarities in the Si or Ca abundances of BH3⋆.

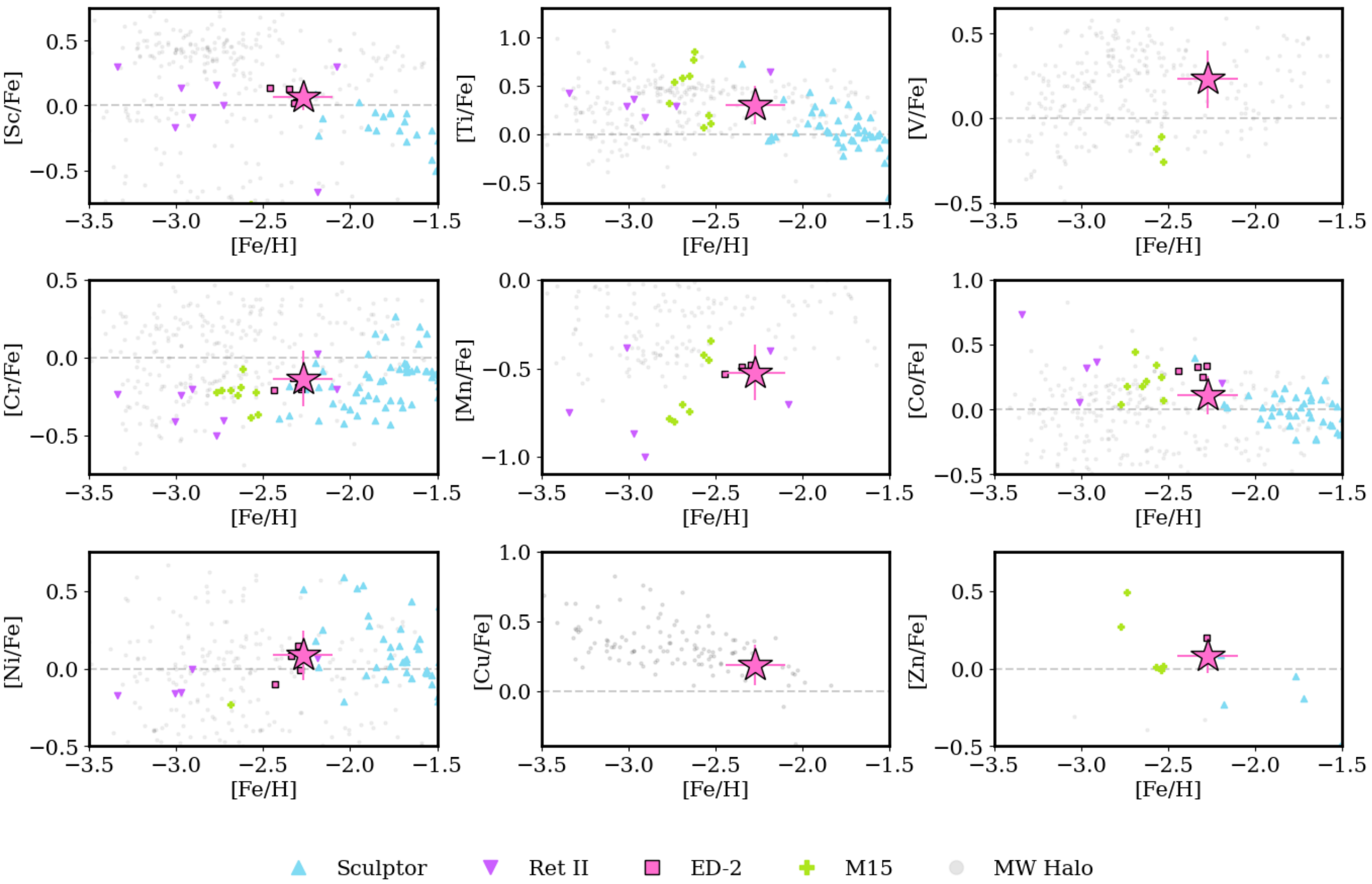

**Figure 5.** Similar to Figure 3 but for Fe-peak elements.

#### *4.2.3. Fe-peak Elements*

We derived the abundances for the 10 Fe-peak elements Sc, Ti, V, Cr, Mn, Fe ,Co, Ni, Cu, and Zn. The Fe-peak abundances for BH3⋆ are plotted in Figure 5 along with Sculptor, Reticulum II, ED-2, M15, and other halo stars.

Type Ia SNe are the primary production sites for Fe-peak elements, occurring when a carbon-oxygen white dwarf accretes matter from a close companion star and exceeds the critical mass to trigger a thermonuclear explosion (B. M. Tinsley 1980; F. Matteucci & S. Recchi 2001; T. C. L. Trueman et al. 2025). There are debates as to whether the companion star is a red giant that overfills its Roche Lobe, causing the white dwarf to exceed its Chandrasekhar mass (J. Whelan & I. Iben 1973) or another white dwarf in which the binary loses angular momentum via gravitational waves (I. Iben & A. V. Tutukov 1984; Z. Han & P. Podsiadlowski 2004). Type II Sne also contribute to iron production (T. Tsujimoto et al. 1995; A. Recio-Blanco et al. 2014), making Fe the prime element for tracing stellar metallicities (e.g., A. McWilliam 1997).

Titanium (Ti) has been previously classified as an $\alpha$ element because it mimics the $\alpha$ elements in the context of Galactic chemical evolution and can be produced by Type II SNe (i.e., K. A. Venn et al. 2004; T. Bensby et al. 2014), yet we classify Ti as an Fe-peak element due its non-negligible contributions from Type Ia SNe (J. D. Keegans et al. 2023; T. C. L. Trueman et al. 2025). The production site of Ti in the Milky Way remains an open question in both observations of SNe and Galactic chemical evolution models (e.g., T. Mishenina et al. 2017; N. Prantzos et al. 2018; C. Kobayashi et al. 2020; T. Sato et al. 2025). We report [Ti I/Fe] $= 0.31 \pm 0.15$ and [Ti II/Fe] $= 0.28 \pm 0.07$. Similar to J. J. Cowan et al. (2020), we find that Ti is enhanced relative to the solar abundance, supporting the notion that Fe-peak synthesis sites produce an overabundance of Ti at low metallicities. The overall Ti abundance of BH3⋆ is shown in the top middle panel of Figure 5, agreeing well with other ED-2 stars and showing no chemical peculiarities.

The odd-*Z* Fe-peak elements scandium (Sc), vanadium (V), manganese (Mn), and cobalt (Co) have similarly debated production sites as Ti. Sc typically follows $\alpha$ elements and is mainly produced in massive stars (F. X. Timmes et al. 1995), while V, Mn, and Co can be produced in both Type II and Type Ia SNe (S. E. Woosley & T. A. Weaver 1995). C. Battistini & T. Bensby (2015) found that Sc, V, and Co loosely track $\alpha$ elements and are likely produced by Type II SNe, while Mn has significant contribution from Type Ia SNe. V, Mn, and Co were estimated each using one line at V II 4875, Mn I 4823, and Co I 5483, respectively. Our Odd-*Z* Fe-peak abundances are [Sc II/Fe] $= 0.07 \pm 0.09$, [V II/Fe] $= 0.28 \pm 0.17$, [Mn I/Fe] $= -0.52 \pm 0.16$, and [Co I/Fe] $= 0.12 \pm 0.12$. It has been shown that Sc, V, and Ti are positively correlated, especially when pushing to stars with [Fe/H] $< -2$ (J. J. Cowan et al. 2020) and our Sc, V, and Ti abundances for BH3⋆ agree with this finding. We show the Sc, V, Mn, and Co abundances in the top left, top right, center, and middle right panels of Figure 5, respectively. There are no obvious chemical peculiarities or departures from other ED-2 abundances in any of these elements.

Cr, Fe, and Ni were largely thought to be produced by Type Ia SNe (F. X. Timmes et al. 1995; S. E. Woosley &

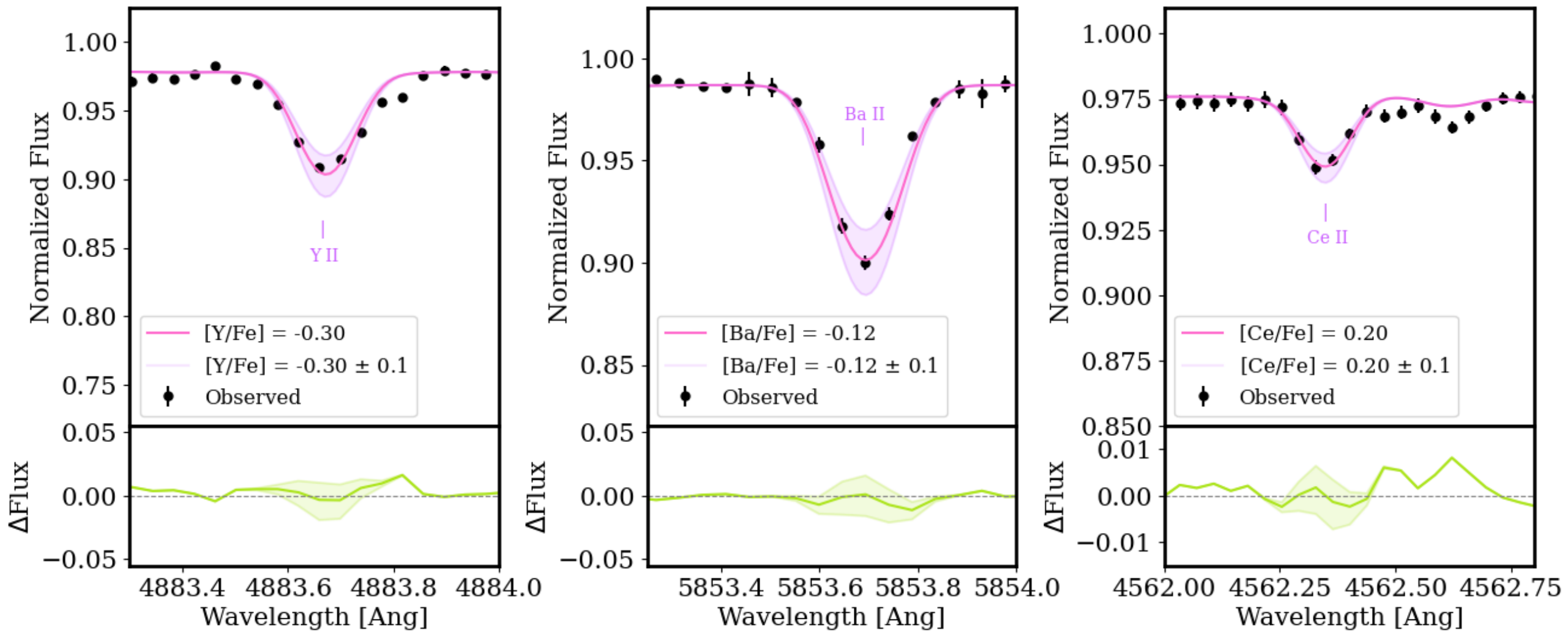

**Figure 6.** We show three lines of different s-process elements, yttrium, barium, and cerium. In all of the panels, the black points are the observed spectrum, the pink lines represent our derived abundances of these respective elements, and the purple region is ±0.1 dex of the abundances. The green regions in the panels below represent the differences between the observed spectrum and the synthetic spectra.

T. A. Weaver 1995), until recently when C. Kobayashi et al. (2020, and references therein) showed that the inclusion of contributions from hypernovae (HNe), a Type II SNe where the progenitor exceeds 30 solar masses, is necessary. Our derived abundances for these elements are [Cr I/Fe] = $-0.20 \pm 0.22$, [Cr II/Fe] = $-0.01 \pm 0.09$, [Fe I/H] = $-2.28 \pm 0.13$, [Fe II/H] = $-2.26 \pm 0.08$, and [Ni I/Fe] = $0.09 \pm 0.12$. The Cr and Ni abundances are shown in the middle left and bottom left panels of Figure 5, respectively. There are no obvious chemical peculiarities in the BH3⋆ abundances of these elements.

Copper and zinc have been shown to have contributions from the *s*-process (e.g., C. Sneden et al. 1991), yet we include them in the Fe-peak section due to the contributions from Type Ia SNe (F. Matteucci & S. Recchi 2001). The production site of Cu is largely dependent on stellar metallicity, with the *s*-process dominating at higher metallicities (S. Bisterzo et al. 2005; M. Pignatari et al. 2010) and massive star explosions dominating at lower metallicities (S. Bisterzo et al. 2004; D. Romano & F. Matteucci 2007). Copper and zinc abundances were derived using the lines Cu I 5153 and Zn I 4810 Å. Our Cu I abundance is [Cu/Fe] = $0.19 \pm 0.14$ dex. The heaviest isotopes of Zn are produced by the *s*-process (S. Bisterzo et al. 2005) and the sources for the lighter isotopes of Zn include SNe, HNe, electron-capture SNe, and $\alpha$-rich freeze-out in massive stars (S. E. Woosley & T. A. Weaver 1995; C. Kobayashi et al. 2006; M. Pignatari et al. 2010; Y. Hirai et al. 2018; E. Caffau et al. 2023). Our Zn I abundance for BH3⋆ is [Zn/Fe] = $0.09 \pm 0.12$. The Cu and Zn abundances are found in the final two panels of Figure 5, illustrating that BH3⋆ has no chemical peculiarities in these elements.

### 4.2.4. Neutron-capture Elements

We attempted to derive abundances of 13 neutron-capture elements in BH3⋆. Of these, 11 elements (Sr, Y, Zr, Ba, La, Ce, Pr, Nd, Sm, Eu, and Dy) were confidently detected. The full list of lines used to derive these abundances is found in Table 4. We find an Eu abundance that classifies this star as a mildly *r*-process enhanced r-I neutron-capture star. (T. C. Beers & N. Christlieb 2005) Figure 6 displays representative synthesis fits for three lines of light neutron-capture elements: Y, Ba, and Ce. The pink line represents our derived average abundances for these elements and the purple region shows a ±0.1 dex region around the line. We show the spectrum of BH3⋆ in black points. The bottom panel shows the differences between our synthesized spectrum and the observed spectrum. Figure 7 plots the abundances of lighter neutron-capture abundances of BH3⋆ compared to other ED-2 stars, along with Sculptor, Reticulum II, M15, and other halo stars.

In neutron-dense environments, the heavier elements on the periodic table are formed through neutron-capture processes, where neutrons will attach to a seed nucleus and eventually decay into a stable element. Elements produced with contributions from the slow neutron capture (*s*-process) range from Sr to Pb. Meanwhile, *s*-process nucleosynthesis typically occurs in AGB stars, in which thermal pulses create a high neutron flux which are then captured by seed nuclei to create *s*-process elements through radioactive decay (R. Gallino et al. 1998; M. Busso et al. 1999; A. I. Karakas & J. C. Lattanzio 2014; G. Contursi et al. 2024). The heaviest elements on the periodic table are synthesized in the rapid neutron-capture process. The astrophysical site(s) for the *r*-process remain uncertain, with proposed sources including neutron star mergers (B. Abbott et al. 2017; M. R. Drout et al. 2017), jet-driven magnetorotational SNe (N. Nishimura et al. 2017), and collapsars (D. M. Siegel et al. 2019). Studies of metal-poor stars enriched in *r*-process material can provide valuable insight into this unresolved question (e.g., D. L. Burris et al. 2000; C. Sneden et al. 2000; V. Hill et al. 2002; C. Sneden et al. 2003; A. P. Ji et al. 2016; A. Bandyopadhyay et al. 2024; I. U. Roederer et al. 2024; S. P. Shah et al. 2024).

Strontium was estimated using one Sr I line at 4305 Å, yielding an abundance of [Sr/Fe] = $0.02 \pm 0.12$. The Sr abundance of BH3⋆ is shown in the top left panel of Figure 7, exhibiting no chemical peculiarities with respect to other ED-2 stars. Y, Zr, Ba, La, Ce, and Nd are all detected using multiple lines. We see no chemical peculiarities in the BH3⋆ abundances of these elements (shown in Figure 7). Praseodymium and samarium were derived using the lines Pr II 4408 Å and Sm II 4577 Å. Although there are no other ED-2 abundances for these elements, the BH3⋆ abundances lie in the range of what is expected compared to the other light neutron-capture element abundances.

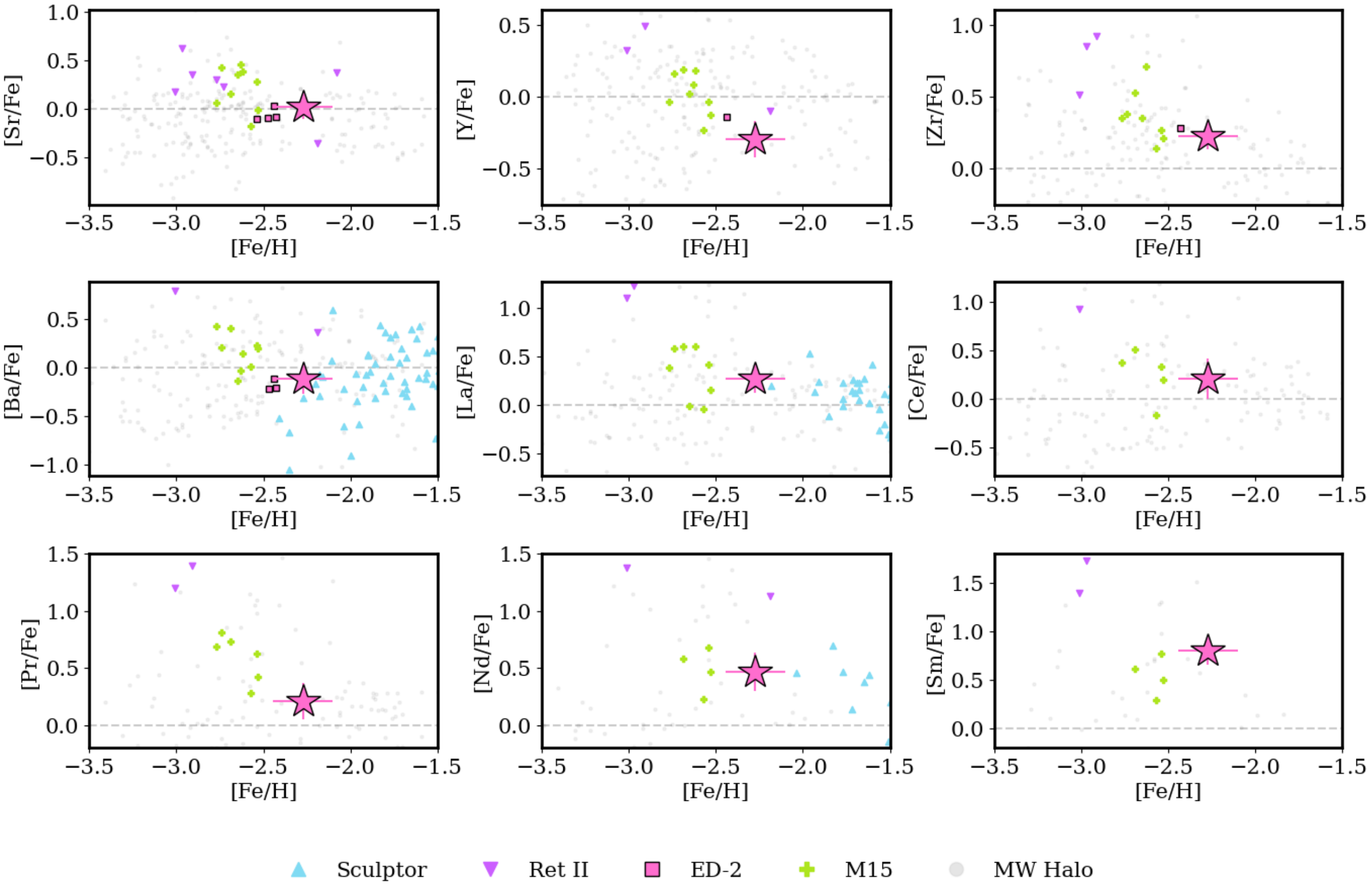

**Figure 7.** Similar to Figure 3 but light neutron-capture elements.

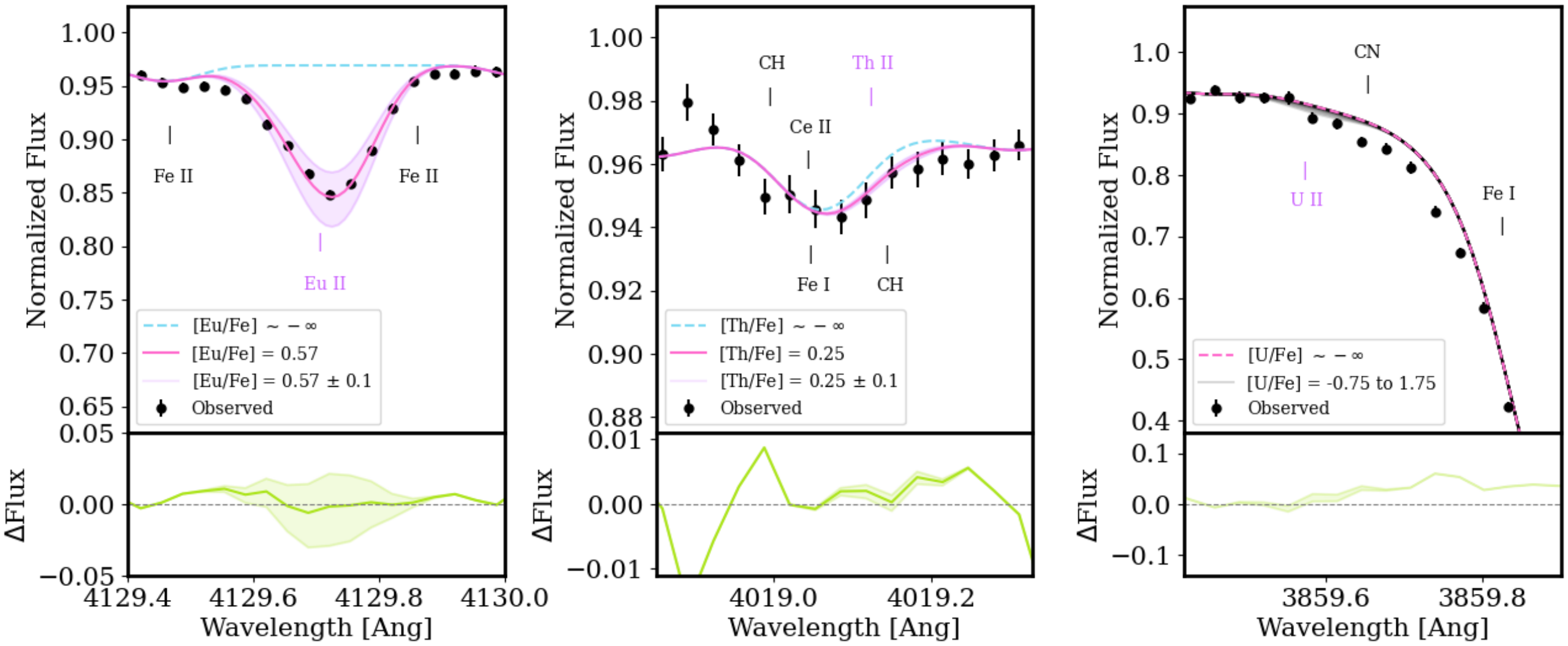

**Figure 8.** Left: The observed spectrum of BH3⋆ in black points, with a synthetic spectrum fit to an Eu abundance of [Eu/Fe] $= 0.57$ in pink and a $\pm 0.1$ dex region around this abundance (purple). Middle: The observed spectrum of the red giant in Gaia BH3 in black points, with a synthetic spectrum fit to a Th abundance of [Th/Fe] $= 0.25$ in pink and a $\pm 0.1$ dex region around this abundance (purple). This highlights the possibility of a Th detection at this line. Right: Observed spectrum (black points) compared to a synthesized spectrum with [U/Fe] $\sim -\infty$ (pink). Gray curves show synthetic spectra from [U/Fe] $= -0.75$ to 1.75 in 0.1 dex steps in the top panel, and the range of differences from the observed spectrum and synthetic spectra are shown in green in the bottom panel. No uranium feature is clearly detected. The nearby Fe I line at 3859.91 Å appears saturated, complicating the measurement of the U II 3859 Å line.

We derived a europium abundance of [Eu/Fe] $= 0.57 \pm 0.19$ using the Eu II 4129 Å line, plotted in the left panel of Figure 8. We find [Dy/Fe] $= 0.67 \pm 0.14$ using the Dy II 4077 Å line. The europium abundance classifies this star as a mildly *r*-process enhanced r-I neutron-capture star (T. C. Beers & N. Christlieb 2005). Due to the lower *r*-process enhancement of BH3⋆, the detection of heavier elements, such as the actinide elements U and Th, is unlikely. Actinide elements are more easily found in stars greatly enhanced in *r*-process elements (i.e., E. M. Holmbeck et al. 2018). Nevertheless, the high SNR of our BH3⋆ spectrum enabled us to search for these lines.

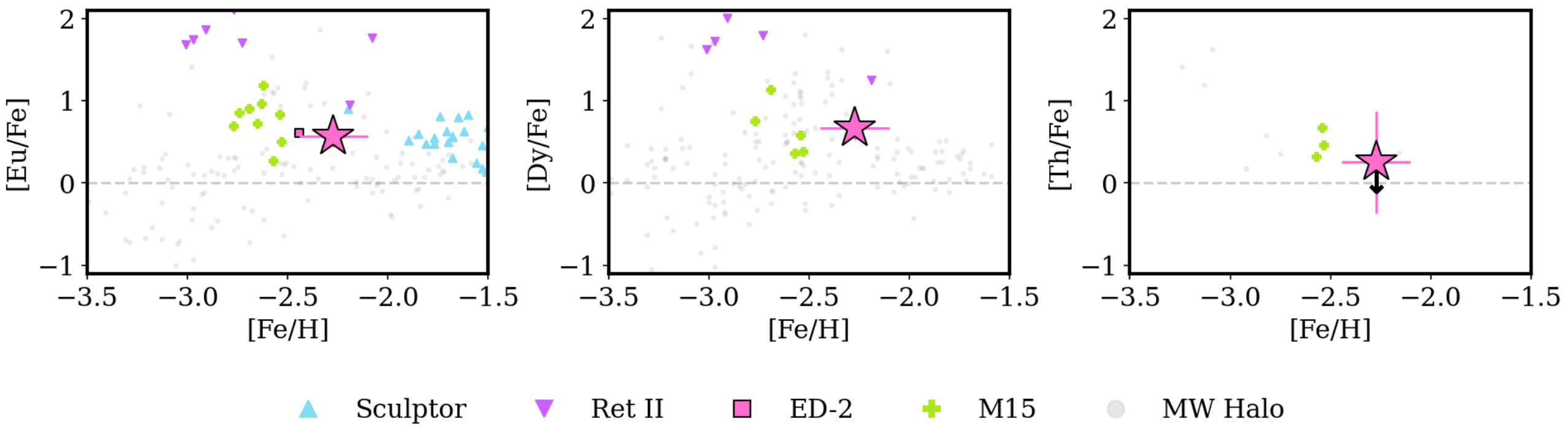

**Figure 9.** Similar to Figure 3 but heavy neutron-capture elements. We show our upper limit detection of Th in the final panel, denoting this abundance with a black arrow.

For Th, which is among the weakest lines we attempt to measure, the automated BACCHUS continuum placement proved unreliable because of the small contrast between the line and the surrounding continuum in this spectral region. Therefore, we manually adjusted the continuum and derived the Th abundance with TURBOSPECTRUM, following the same synthesis procedure as in BACCHUS. There is a probable Th detection at the Th II 4019 Å line, shown in the middle panel of Figure 8. We show that there is a distinction between "no thorium" ([Th/Fe] $= -\infty$, blue dashed line) and a [Th/Fe] abundance of 0.25, and consequently adopt this abundance as an upper limit. The abundances of the heavy neutron-capture elements Eu, Dy, and Th for BH3⋆ are plotted in Figure 9 along with Sculptor, Reticulum II, ED-2, M15, and other halo stars. There are no chemical peculiarities in these heavy *r*-process elements, though this interpretation is limited by the small sample size of other stars with confident heavy *r*-process element detections.

We searched for uranium in BH3⋆ using the U II 3859 Å line, as shown in the right panel of Figure 8. This line is heavily blended with CN features and lies in the wing of a strong Fe I line, making it extremely challenging to measure. For the CN feature, we fix our C abundance to our derived abundance of [C/Fe] = 0.27 and vary the N abundance to extremes ($\pm$2.0 dex), starting at the BH3⋆ N abundance of [N/Fe] = −0.40 derived in E. Dodd et al. (2025) to quantify if this had an effect on the detectability of the U line, but found no significant effect.

In the right panel of Figure 8, we present synthetic spectra with [U/Fe] values ranging from -0.75 to 1.75 dex in 0.1 dex steps (gray lines), along with a "no uranium" model ([U/Fe] $= -\infty$, pink dashed line). The lack of any significant deviation between the synthetic spectra and the no uranium case leads us to conclude that uranium is not detected in this star. We looked for U detections at the 4050 and 4090 Å lines proposed by S. P. Shah et al. (2023) but found no evidence of uranium due to the intrinsic weakness of these lines and the lower SNR in this spectral region.

With our abundances of multiple neutron-capture elements, we compare the *r*-process abundance pattern of BH3⋆ to the solar *r*-process abundance pattern. Stars across a range of metallicities exhibit a homogeneous pattern in the elemental abundances of *r*-process elements (C. Sneden et al. 2002; A. P. Ji et al. 2016; T. T. Hansen et al. 2017; X.-J. Xie et al.

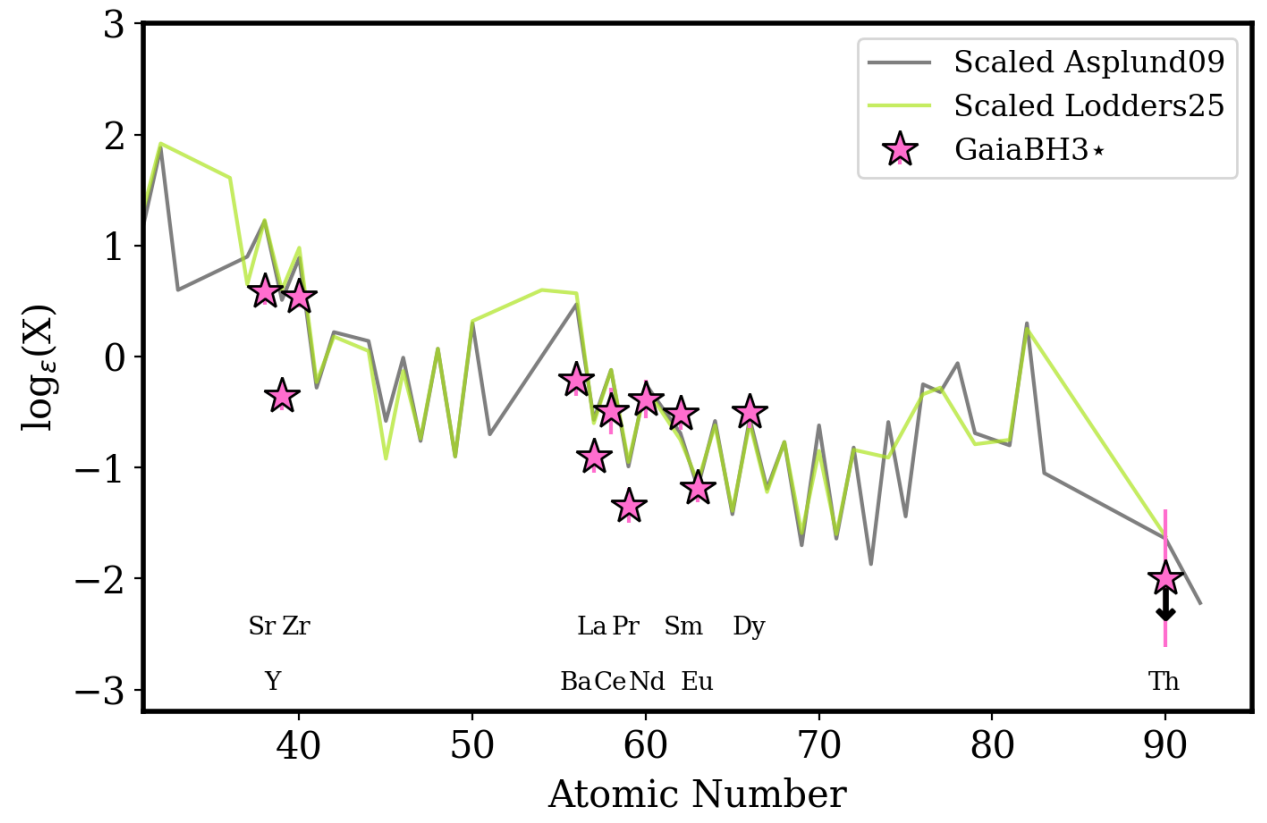

**Figure 10.** The *r*-process pattern plot for BH3⋆ (pink stars). Two scaled solar abundances are shown in gray (M. Asplund et al. 2009) and green (K. Lodders et al. 2025). The abundances from BH3⋆ generally follow the universal pattern of the *r*-process and thorium is denoted as an upper limit with a black arrow.

2025), suggesting that the *r*-process yields are largely uniform across different Galactic environments (e.g., Milky Way halo stars, ultrafaint dwarfs, and classical dSphs). We scale two solar *r*-process patterns (M. Asplund et al. 2009; K. Lodders et al. 2025) to our Eu abundance of $A$(Eu) = −1.70 and show the *r*-process pattern of BH3⋆ in Figure 10. While the *r*-process abundances of BH3⋆ tend to match the solar pattern, it is unsurprising that there some offsets. A. Bandyopadhyay et al. (2025) found large intrinsic scatters in the *r*-process pattern of the GC NGC 2298. This illustrates that while the *r*-process pattern could be considered universal, environments such as GCs can show large *r*-process dispersions among the stellar populations, hinting at inhomogeneous mixing or multiple *r*-process nucleosynthetic sites.

Although the neutron-capture abundances of BH3⋆ are consistent with those of other ED-2 stars and follow the general expected *r*-process pattern, its classification as an r-I star is noteworthy. P. S. Barklem et al. (2005) found that only about 15% of halo stars fall into the r-I category, making such stars relatively rare. Given the strong diagnostic power of *r*-process elements in tracing nucleosynthetic pathways and Galactic enrichment processes (i.e., T. Hansen et al. 2011; T. Tsujimoto & T. Shigeyama 2014; C. Manea et al. 2024; S. Monty et al. 2024), the *r*-process signature in BH3⋆ can provide valuable constraints on its formation history and the broader chemical evolution of its birth environment.

### 4.3. Cosmochronometry of BH3⋆

One of the benefits of characterizing the neutron-capture elements of BH3⋆ is the possibility of placing a nuclear cosmochronometric age on this star. Using the fact that elements such as Th and U decay on cosmological timescales, we can adopt heavy-element nuclear chronometers to estimate the age of the gas from which the star formed (J. J. Cowan et al. 1991; C. Sneden et al. 2002; V. M. Placco et al. 2017).

If Gaia BH3 formed and evolved through isolated binary evolution, placing an age on BH3⋆ would place an age on the black hole. Asteroseismology is widely considered the gold standard for determining stellar ages (A. Miglio et al. 2013), but no asteroseismic signal has been detected in the NASA Transiting Exoplanet Survey Satellite (G. R. Ricker et al. 2014) data for BH3⋆ (D. Hey et al. 2025). Thus, we attempted to estimate the star's age using nuclear cosmochronometry. Heavy elements such as U and Th decay on cosmological timescales, with half-lives of ∼4.47 Gyr and ∼14 Gyr, respectively. We can use the detections of these elements in stars to age the *r*-process enriched gas from which stars formed (J. J. Cowan et al. 1991; C. Sneden et al. 2003; A. Frebel et al. 2007; V. Hill et al. 2017; S. P. Shah et al. 2023; Y. Y. Huang et al. 2025). The most precise chronometer pair would be U/Th, but we are limited to Th/Eu. For this exercise, we assumed that our Th upper limit is a fixed Th abundance of [Th/Fe] = 0.25 and used the following equation from V. Hill et al. (2002):

$$\tau = 46.7\,\mathrm{Gyr}\,[\log(\mathrm{Th/Eu})_{\mathrm{o}} - \log(\mathrm{Th/Eu})_{*}]. \quad (1)$$

We adopted the initial production ratio ($\log(\mathrm{Th/Eu})_{\mathrm{o}}$) from H. Schatz et al. (2002) where these ratios were found with waiting-point calculations. The Th and Eu abundances of BH3⋆ produced a tentative age of 22.8 ± 5.6 Gyr. This age is improbable because it is larger than the current estimated age of the Universe, yet we note that our observational errors on Th are large and this method provides a rough upper limit on the age of the gas from which BH3⋆ formed. Another reason for this unphysical age is that the errors on nuclear cosmochronometric ages are still quite large, with dominant error sources being the initial production ratios and chosen chronometer pairs. The Th/Eu chronometer has been shown to be highly sensitive to the chosen production ratio and *r*-process site (V. Hill et al. 2017; E. M. Holmbeck et al. 2018; A. Azhari et al. 2025). The U/Th chronometer can reduce the systematics in this equation, thus a detection of U in this star would greatly assist in narrowing down the nuclear cosmochronometric age range of BH3⋆.

Using an isochrone fit to ED-2 stars, E. Dodd et al. (2025) concluded an age for ED-2 of 13.4 ± 1.6 Gyr. E. Balbinot et al. (2024) found that ED-2 stars match the color–magnitude diagram of M92, an old metal-poor GC with an age of 13.80 ± 0.75 Gyr (J. M. Ying et al. 2023). With our tentative upper limit on [Th/Fe] and the known challenges in using the Th/Eu chronometer, we do not yet find a nuclear cosmochronometric age consistent with these age determinations. When fixing the age to the isochrone age of 13.4 Gyr, the production ratio from H. Schatz et al. (2002), and our Eu abundance, the thorium abundance required to produce the age is about 1 dex higher than our derived Th abundance at $\log_\epsilon(\mathrm{Th})$ = 1.14. We show this fixed abundance in Figure 11, illustrating that this abundance lies multiple sigmas away from what is observed in BH3⋆. If we instead fix our thorium abundance and vary the production ratio, we find $\log(\mathrm{Th/Eu})_{\mathrm{o}} = -0.533$ which is over 0.2 dex lower than different production ratios determined by H. Schatz et al. (2002), V. Hill et al. (2002), and I. U. Roederer et al. (2009).

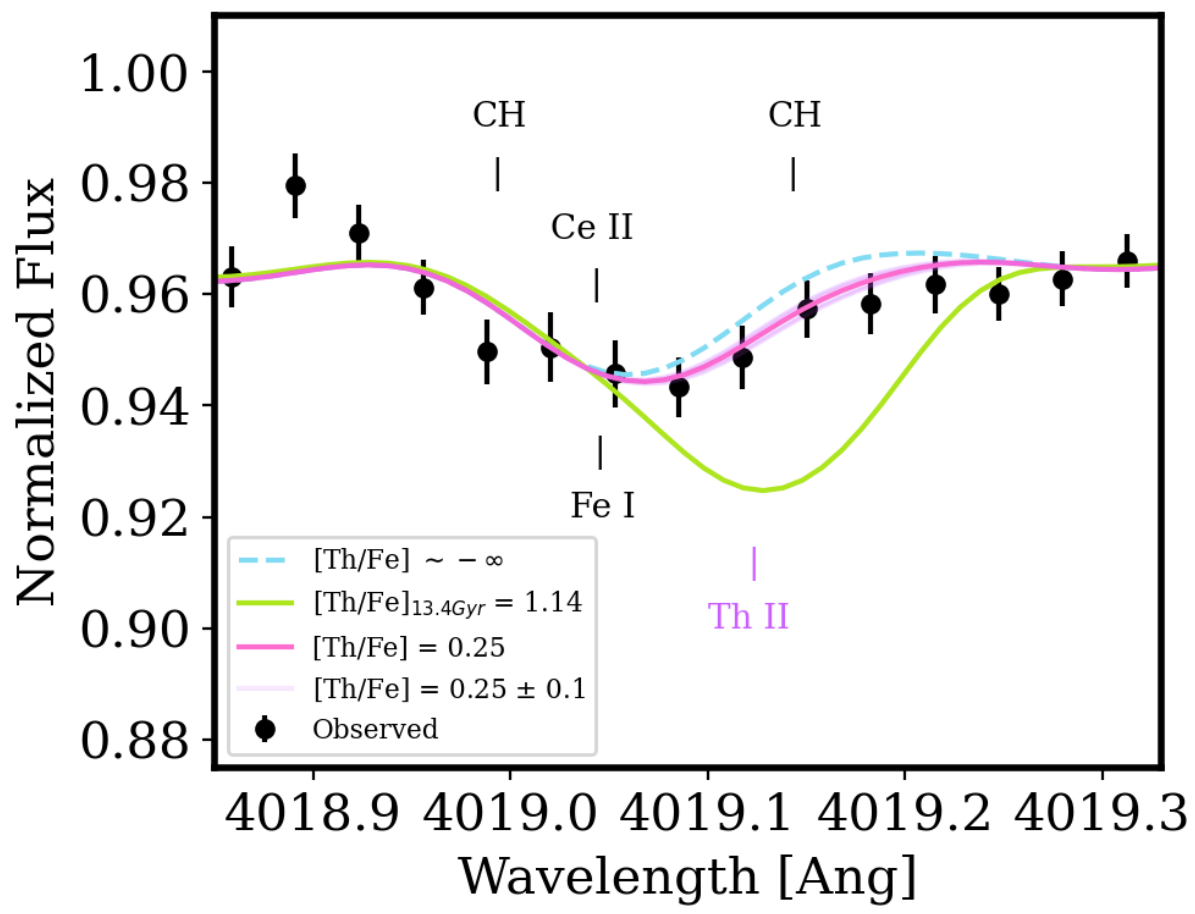

**Figure 11.** Similar to the middle panel of Figure 8, we show the observed spectrum of BH3⋆ in black points with a synthetic spectrum fit to [Th/Fe] = 0.25 (pink) and a ±0.1 dex region around our fit (purple). A Th abundance of [Th/Fe]= −∞ is shown with the blue dashed line. We then fix an age of 13.4 Gyr and the H. Schatz et al. (2002) production ratio to solve for the [Th/Fe] needed, resulting in [Th/Fe] = 1.14 (green). We show that the Th abundance found using the probable age and production ratio cannot conceivably fit our observed data.

A. P. Ji & A. Frebel (2018) found a similarly unphysical age when applying the Th/Eu chronometer to Reticulum II stars, with the youngest age coming out to be ∼21.7 Gyr. The authors concluded that the current production ratios do not describe the *r*-process evolution of Ret II, and we suggest that this limitation may also apply to the progenitor evolution of ED-2. Although our age result for BH3⋆ does not align with other age determinations, it highlights both the promise and the current limitations of nuclear cosmochronometry as an age-dating technique.

## 5. Conclusions

Gaia BH3 is the third black hole discovered with Gaia astrometry—a 33 $M_\odot$ mass black hole in a binary system with a red giant star, located in the halo stream ED-2 (E. Balbinot et al. 2024). This red giant (BH3⋆) is old, metal-poor ([Fe/H] = −2.27 ± 0.15), $\alpha$-enhanced ([$\alpha$/Fe] = 0.41), and *r*-process enhanced ([Eu/Fe] = 0.57 ± 0.19). The formation pathway of this system is still debated, with the two likely candidates being isolated binary evolution or dynamical capture. In both cases, chemical peculiarities of BH3⋆ are predicted to be unlikely. Here, we present the most comprehensive chemical analysis to date of the red giant in Gaia BH3 and aim to confirm the lack of chemical peculiarities.

We obtained 43.5 hr of high-resolution spectra of BH3⋆ from the Tull Coudé spectrograph on the 2.7 m Harlan J. Smith Telescope at McDonald Observatory to characterize the abundances of BH3⋆. We determined the stellar parameters of BH3⋆ photometrically, yielding $T_{\mathrm{eff}}$ = 5416 ± 84 K and log $g$ = 3.00 ± 0.14 using the $G$-band magnitude from Gaia Collaboration et al. (2022) and the $K_s$ magnitude from 2MASS (R. M. Cutri et al. 2003). Employing the stellar parameter and chemical abundance code BACCHUS, we derived the abundances of 29 elements in BH3⋆, ranging from lithium to thorium.

Of these elements, we show that BH3⋆ has an expected Li abundance of a red giant star and lies below the primordial Li abundance illustrated by the Spite plateau (F. Spite & M. Spite 1982) with $A(\mathrm{Li}) = 1.34$ dex. This star is C-normal with a carbon abundance of $[\mathrm{C/Fe}] = 0.27 \pm 0.19$. The $\alpha$ abundances of O, Mg, Si, and Ca show that this star is $\alpha$-enhanced, typical of a halo giant. We find no chemical peculiarities in the nine Fe-peak abundances of BH3⋆. There are neutron-capture elements confidently detected in this star ranging from Sr to Dy, with abundances typical of an r-I neutron-capture star. In our full suite of derived abundances, there are no chemical peculiarities of BH3⋆, even when compared to other ED-2 stars (E. Dodd et al. 2025). The heavier elements of this star generally follow the universal *r*-process abundance pattern when compared to the scaled solar pattern.

Looking toward the actinide elements U and Th, we find no confident U detection in this star. We find a tentative detection of Th using the Th II 4019 Å, resulting in an upper limit of $[\mathrm{Th/Fe}] \lesssim 0.25 \pm 0.64$. If we assume that the [Th/Fe] abundance is a detection, we can use the Th/Eu chronometer to attempt to place an age on this star. This resulted in an upper limit age of $22.8 \pm 5.6$ Gyr, which is unphysical when comparing to the estimated age of the Universe. A discussion of other age determinations and the limitations of nuclear cosmochronometry can be found in Section 4.3.

With the most comprehensive list of chemical abundances to date, we find no chemical peculiarities in BH3⋆. The chemical "normalcy" of this star is consistent with both formation theories for Gaia BH3, including dynamical capture and isolated binary evolution. Although this system represents one of the first known black hole binaries with a low-metallicity luminous companion discovered with Gaia astrometry, many more long-period systems of this kind are anticipated to be discovered with the forthcoming release of Gaia DR4 (K. El-Badry 2024; Gaia Collaboration et al. 2024). Efforts to identify similar systems are already underway (e.g., P. Nagarajan et al. 2025; J. Müller-Horn et al. 2025) and this work will serve as a foundation for future chemical analyses of luminous companions of black holes revealed in Gaia DR4.

## Acknowledgments

We thank the referee for their insightful comments. Z.H. thanks Chris Sneden for invaluable help and insight on this work. Z.H. thanks Amaya Sinha and Natalie Myers for countless BACCHUS discussions. Z.H. additionally thanks all staff and observing support at McDonald Observatory.

K.H. and Z.H. are partially supported by NSF AST-2407975. K.H. acknowledge support from the Wootton Center for Astrophysical Plasma Properties, a U.S. Department of Energy NNSA Stewardship Science Academic Alliance Center of Excellence supported under award numbers DE-NA0003843 and DE-NA0004149, from the United States Department of Energy under grant DE-SC0010623. This work was performed in part at the Aspen Center for Physics, which is supported by National Science Foundation grant PHY-2210452. C.M. is supported by the NSF Astronomy and Astrophysics Fellowship award No. AST-2401638.

This work has made use of data from the European Space Agency (ESA) mission Gaia (https://www.cosmos.esa.int/gaia), processed by the Gaia Data Processing and Analysis Consortium (DPAC, https://www.cosmos.esa.int/web/gaia/dpac/consortium). Funding for the DPAC has been provided by national institutions, in particular the institutions participating in the Gaia Multilateral Agreement.

*Facilities:* Smith (Tull Coudé Spectrograph).

*Software:* `astropy` (Astropy Collaboration et al. 2022, 2013), `BACCHUS` (T. Masseron et al. 2016), `iSpec` (S. Blanco-Cuaresma et al. 2014; S. Blanco-Cuaresma 2019), `matplotlib.pyplot` (J. D. Hunter 2007), `numpy` (S. van der Walt et al. 2011), `scipy` (P. Virtanen et al. 2020), `TSDRP`,[5] `TURBOSPECTRUM` (B. Plez 2012).

## ORCID iDs

Zoe Hackshaw https://orcid.org/0000-0002-3855-3060
Keith Hawkins https://orcid.org/0000-0002-1423-2174
Catherine Manea https://orcid.org/0000-0002-0900-6076

[5] https://github.com/grzeimann/TSDRP